\documentclass{ws-jai}
\usepackage[flushleft]{threeparttable}
\begin{document}

\catchline{}{}{}{}{} 

\markboth{Izumi Mizuno}{Software Polarization Spectrometer ``PolariS''}

\title{Software Polarization Spectrometer ``PolariS''}

\author{Izumi Mizuno$^{1, 2}$, Seiji Kameno$^3$, Amane Kano$^4$, Makoto Kuroo$^{5}$, Fumitaka Nakamura$^6$, Noriyuki Kawaguchi$^6$, Katsunori M. Shibata$^6$, Seisuke Kuji$^6$, and Nario Kuno$^2$}

\address{
$^1$Department of Physics and Astronomy, Graduate School of Science and Engineering, Kagoshima University, 1-21-35 Korimoto, Kagoshima, 890-0065 Japan; k7330312@kadai.jp \\
$^2$Nobeyama Radio Observatory, Minamimaki, Minamisaku, Nagano 384-1805 Japan,\\
$^3$Joint ALMA Observatory, Alonso de C\'ordova 3107 Vitacura, Santiago 7630355 Chile\\
$^4$Department of Physics, Faculty of Science, Kagoshima University, 1-21-35 Korimoto, Kagoshima, 890-0065 Japan \\
$^5$Shoyo High School, Tatemachi, Hachioji, Tokyo, 293-0944 Japan \\
$^6$National Astronomical Observatory of Japan, 2-21-1 Osawa, Mitaka, Tokyo, 181-8588 Japan \\
}

\maketitle

\begin{history}
\received{(to be inserted by publisher)};
\revised{(to be inserted by publisher)};
\accepted{(to be inserted by publisher)};
\end{history}

\begin{abstract}
We have developed a software-based polarization spectrometer, PolariS, to acquire full-Stokes spectra with a very high spectral resolution of 61 Hz.
The primary aim of PolariS
is to measure the magnetic fields in dense star-forming cores
by detecting the Zeeman splitting of molecular emission lines.
The spectrometer
consists of a commercially available
digital sampler
and a Linux computer.
The computer is equipped with a
graphics processing unit (GPU) to process
FFT and cross-correlation using the CUDA (Compute Unified Device Architecture) library developed by NVIDIA.
Thanks to a high degree of precision in quantization of the analog-to-digital converter and arithmetic in the GPU, PolariS offers excellent performances in linearity, dynamic range, sensitivity, bandpass flatness and stability.
The software has been released under the MIT License and is available to the public.
In this paper, we report the design of PolariS and its performance verified through engineering tests and commissioning observations.

\end{abstract}


\section{Introduction}

\subsection{Polarization spectrometer}
A polarization spectrometer is a device to measure the polarimetric properties of incident radiation as a function of frequency.
The Stokes parameters, $I$, $Q$, $U$ and $V$, are a representation of polarized light, and are all measurements of flux density.
The total flux density 
is denoted by $I$, and the level of general elliptical polarization is described by nonzero values of $Q$, $U$ and $V$. 

A typical astronomical receiving system is sensitive to either linear or circular polarization. In both cases, the receiving system will measure two products, either two orthogonal linear polarizations or two opposing circular polarizations. In order to obtain all four Stokes parameters from the two observables, both auto- and cross-correlation products must be formed from the received polarizations.

In a modern radio telescope, received signals are numerically processed to output correlation products in a digital polarization spectrometer.
The XPOL correlation polarimeter system is installed on the IRAM 30-m telescope \cite{2008PASP..120..777T} and the VESPA digital polarization spectrometer of the system yields a 40-kHz spectral resolution across a 120-MHz bandwidth.
The ACS spectrometer is installed on the Green Bank Telescope and offers a 49-kHz spectral resolution in a 12.5-MHz bandwidth.
The WAPP correlator \cite{2000ASPC..202..275D} is used in the Arecibo radio telescope mainly for pulsar observations, with a 0.195 MHz resolution.
Such hardware-based polarization spectrometers are not sufficiently flexible nor scalable for finer spectral resolutions, wider bandwidths, etc.

%
A software-based polarization spectrometer, composed by a personal computer, is an alternative implementation to bring flexibility and scalability.
Recent high-performance computing system with standard off-the-shelf Graphics Processing Units (GPUs) provides solutions for real-time spectroscopy (e.g. \cite{2011MNRAS.417.2642M, 2012MNRAS.422..379B}) with high-level programming interfaces such as CUDA \cite{2007NVIDIA}.
The software implementation yields some advantages.
It is flexible and scalable when modification of specifications are required.
The source codes in programming language are readable, verifiable, and can be shared or updated via a public repository.
The processors are off-the-shelf products, which allows us to build copies or to prepare spares at a reasonable cost.

In this paper, we describe the construction of the software-based polarization spectrometer, PolariS.
The scientific requirements and specifications to meet are described in section \ref{science_target}.
The basic design and costs are presented in section \ref{design}.
The verification is reported in section \ref{sec:verification}.
And in section \ref{sec:fieldtest}, field test results of observations toward radio sources are demonstrated.

\subsection{Science target}\label{science_target}
%
 The primary aim of PolariS is to measure the strength of the magnetic fields in the star-forming cores of molecular clouds through the Zeeman effect of emission lines.
The magnetic field is a key control parameter that can mediate gravitational collapse of these cores \cite{spitzer1968}, and the Zeeman effect is one of the most reliable probes
to measure the strength of magnetic fields. While magnetic field strength in the interstellar medium has been measured toward diffuse and dark clouds with {\sc Hi} and OH \cite{1986ApJ...301..339T, 1989ApJ...338L..61G, 1993ApJ...407..175C}, star-forming regions with OH masers \cite{1975ApJ...202..650L, 1980ApJ...239...89R, 1987ApJS...65..193G}, OH outflows \cite{1984MNRAS.207..127N, 1986MNRAS.219..145B}, interstellar H$_2$O maser clumps \cite{1989A&A...214..333F}, and massive star-forming regions with CN radical \cite{1999ApJ...514L.121C},
those in the star-forming cores have not been measured to date.

According to the relation of gas density and magnetic field strength \cite{1989A&A...214..333F}, the field strength in the star-forming cores with the gas density of $n_{\rm H} \sim 10^4 - 10^5$ cm$^{-3}$ is expected to lie between 100 and $1000 \ \mu$G.
With typical parameters of the core radius of $\sim 0.1$ pc and the mass of $\sim 10 \ {\rm M}_{\odot}$ \cite{2002ApJ...575..950O}, the gravity-equivalent magnetic field strength will be $\sim 100 \ \mu$G \cite{Shu87}.
Thus, measurements of $\sim 100 \ \mu$G magnetic fields in star-forming cores are crucial to discriminate whether the magnetic fields can support a core against gravitational collapse.
The CCS radical is one of the best molecules to investigate magnetic field strength in star-forming cores because it is abundant in dark cloud cores, its emission lines are intense and narrow, and it shows a relatively large Zeeman split with $0.64\pm0.12$ Hz $\mu$G$^{-1}$ for the $J_N = 4_3 - 3_2$ transition at 45.379 GHz \cite{shinnaga00}.

\subsection{Requirements}\label{Requirements for the Zeeman split}
The specification of PolariS must be compliant to measure the Zeeman split of the CCS molecular emission.
The expected magnetic field of $\sim 100 \ \mu$G will yield the Zeeman
split
of $\sim 64$ Hz \cite{shinnaga00} between two orthogonal circular polarizations.
The Zeeman split will appear in the Stokes $V$ (circular polarization) spectrum as
\begin{eqnarray}
V(\nu) \simeq \frac{dI(\nu)}{d\nu} \Delta \nu, \label{eqn:StokesV}
\end{eqnarray}
where $I(\nu)$ is the Stokes $I$ spectrum as a function of the frequency, $\nu$, and $\Delta \nu$ is the Zeeman split.
Our pilot observations of CCS molecules (see section \ref{sec:fieldtest}) indicate that the line profile toward the Taurus molecular cloud 1 has $\displaystyle \frac{dI}{d\nu} \sim 0.16$ mK Hz$^{-1}$, which corresponds to $V \sim 10$ mK
if we assume that $\Delta \nu = 64$ Hz.

In our project we use the Z45 receiver \cite{2013ASPC..476..403T, 2013ASPC..476..239N}, which is capable of receiving two orthogonal linear polarizations.
The Stokes $V$ component can be obtained via cross correlation between two linear polarizations.
To measure such a weak and narrow Stokes $V$ feature with Z45,
PolariS must specify (1) measurements of autocorrelation and cross correlation, (2) the spectral resolution better than the expected Zeeman
split of 64 Hz, (3) adequate sensitivity, and (4) frequency stability better than 60 Hz.
 The total cost was limited to our budget of JPY 2,000,000.

\section{Design and Development}\label{design}
\begin{figure}[ht]
\begin{center}
\includegraphics[width=160mm]{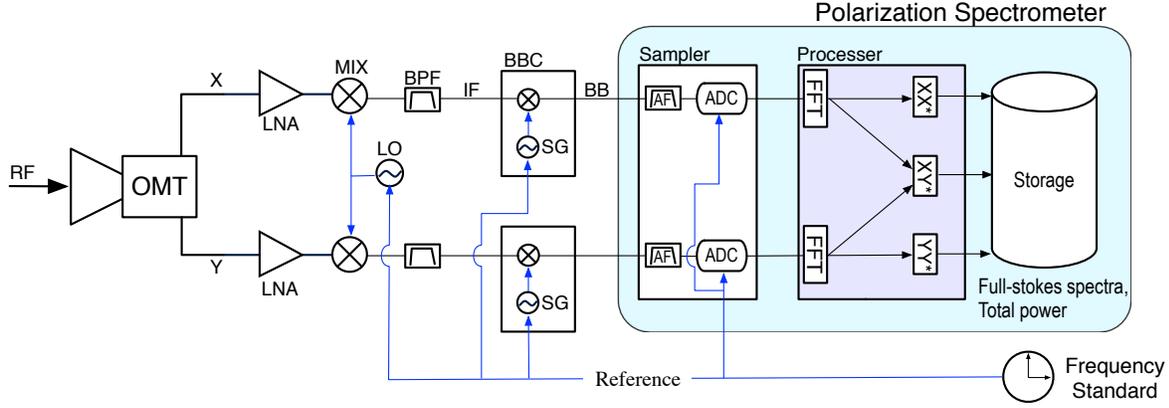}
\end{center}
\caption{
Schematic diagram of the signal flow of the polarization detection system. 
The RF (radio frequency) waves incident on the dual polarization receiver will be split into two polarizations ($X$ and $Y$) in the OMT (ortho-mode transducer), 
amplified in the LNAs (low noise amplifiers), and downconverted in the MIXs (mixers) using the LO (local oscillator) signal referencing the frequency standard. 
The IF (intermediate frequency) signals through BPFs (bandpass filters) will be downconverted again to basebands (BBs) with BBCs (baseband converters) 
at desired frequency of SGs (signal generator), and input to a polarization spectrometer. 
The polarization spectrometer in pale blue consists of a sampler, a processor in pale violet, and storage. The sampler consists of AFs (anti-alias filters) and
ADCs. 
The digitized signals are converted into spectra though
an FFT process, 
and then multiplications between them are taken to output power and cross spectra ($XX^*$, $XY^*$, and $YY^*$), which yield full Stokes spectra. 
Detailed signal processing inside the spectrometer is described in Fig.~\ref{fig:polariS_Inside}.
}
\label{fig:polariS_Connection}
\end{figure}

Fig.~\ref{fig:polariS_Connection} shows a schematic diagram of the signal path to a polarization spectrometer.
The functions of the digital FX-type polarization spectrometer are (1) to accept the received signal in two orthogonal polarizations in digital, (2) to calculate spectra of the real-time signal streams via
FFT, (3) to produce power spectra and cross power spectra by multiplying the spectra, and (4) to integrate the power spectra and output them.

\begin{wstable}[h]
\caption{Specifications of PolariS}
\begin{tabular}{@{}llll@{}} \toprule
Feature & \multicolumn{2}{c}{Specification} & Remarks \\ \colrule
Number of IFs & \multicolumn{2}{l}{4 ($X_0, X_1, Y_0, Y_1$)\tnote{a}} & 2 pols., 2 IFs \\
Bandwidth     & 8 MHz / IF   & 4 MHz / IF & 16 or 8 Msps sampling \\
Quantization  & 4 bit/sample & 8 bit/sample & \\
Spectral resolution & 131072 ch/IF & 65536 ch/IF & 61-Hz channel spacing \\
Autocorrelation & \multicolumn{2}{l}{4 real products} & $\left<X_0X^*_0\right>$, $\left<X_1X^*_1\right>$, $\left<Y_0Y^*_0\right>$, $\left<Y_1Y^*_1\right>$ \\
Cross-correlation & \multicolumn{2}{l}{2 complex products} & $\left<X_0Y^*_0\right>$, $\left<X_1Y^*_1\right>$\\
Time resolution & \multicolumn{2}{l}{1 sec} \\ \botrule
\end{tabular}
\begin{tablenotes}
\item[a] $X, Y$ stands for two orthogonal polarizations. Suffix indicates basebands.
\end{tablenotes}
\label{tab:spec}
\end{wstable}
Table \ref{tab:spec} lists the specifications of PolariS.
The specification of the 61-Hz resolution at 4 or 8 MHz bandwidth requires 65536 or 131072 spectral channels.
To produce the spectrum in real time, 262144- or 131072-point FFT must be accomplished within less than 8 msec.
This processing requires 1440 Mflops per IF or 5600 Mflops for all IFs.
Adding 262 Mflops for autocorrelation cross products, at least $\sim 6$ Gflops are necessary for real-time processing.

\begin{figure}[ht]
\begin{center}
\includegraphics[width=160mm]{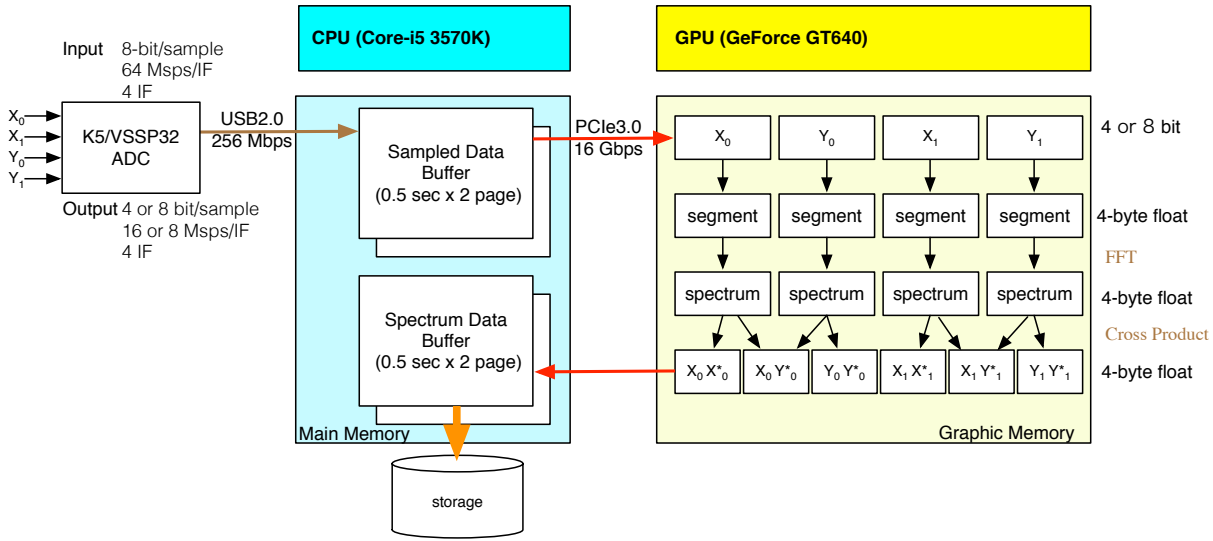} 
\end{center}
\caption{Schematic diagram of
the system and signal flow inside. PolariS consists of the K5/VSSP32
digital sampler
and the host computer including CPU in pale blue, GPU in yellow, and storage.
The sampler consists of four analog-to-digital converters (ADCs) and
accepts four analog IF signals of 32-MHz bandwidth and digitizes them with a sampling speed of 64 Msps/IF and quantization of 8 bit. The digitized signal will be output with 4-bit quantization when we need 8-MHz bandwidth (16 Msps/IF) after applying digital filtering to reduce the total data rate down to 256 Mbps to meet USB2.0. The CPU captures 32 MBytes of data twice a second in the two-page buffer and sends them to GPU memory. The GPU processes data translation (4 or 8-bit integer to 4-byte float), segmentation, FFT, cross products, and accumulation. Each segment contains 131072 or 262144 samples to produce 65536 or 131072 spectral channels. The (cross) power spectra will be transmitted to the main memory and stored in the file system.}
\label{fig:polariS_Inside}
\end{figure}

To meet the required specifications within the limited
budget, we decided to design PolariS as a software-based spectrometer that consists of a VLBI digitizer, K5/VSSP32 \cite{Kondo2006b} for data acquisition, and
a general-purpose personal computer with a graphic processing unit (GPU) for signal processing.
Thanks to recent advances in digital technology,
personal computers have become capable of wide-bandwidth signal processing.
Furthermore, recent GPUs for video games perform TFLOPS (tela floating operations per second) arithmetic with a low cost.

Fig.~\ref{fig:polariS_Inside} shows the signal processes in PolariS. 
 The K5/VSSP32 VLBI sampler accepts four analog signals at a maximum bandwidth of 32 MHz per input. After digitization with 8-bit quantization, we employ digital filtering and quantization level reduction \cite{Kondo2006a} in order to reduce the output data rate to 256 Mbps to accommodate the USB 2.0 interface.
The current PolariS system allows two modes: (1) 8-MHz (16 Msps) bandwidth, 4-bit quantization, and (2) 4-MHz bandwidth, 8-bit quantization.

 The host computer of PolariS works on the Linux Ubuntu 12.04 LTS operating system. The control and signal-processing software are coded in GCC (GNU compiler collection) and NVCC (NVIDIA CUDA compiler). The cuFFT library is employed
o transform time-domain signals into frequency-domain spectra. Total power of each IF signal is measured by histograms of digitized signals \cite{2010PASJ...62.1361N}. All of the source code is available in the GitHub\footnote{https://github.com/kamenoseiji/PolariS} repository. The computer is equipped with the Intel Core-i5 CPU, 4-GB memory, NVIDIA GT640 GPU. It costs JPY 70,000.

\subsection{Measurement of the Stokes parameters} \label{sec:stokes}
PolariS produces power spectra of $\left<XX^*\right>$ and $\left<YY^*\right>$ and cross power spectra of $\left<XY^*\right>$. Full Stokes spectra of $I$, $Q$, $U$, and $V$ are derived using
these products offline.

If $X$ and $Y$ stand for linear polarizations without any cross talk between them, the Stokes parameters are derived using the equation:
\begin{eqnarray}
\left( \begin{array}{c}
I \\
Q \\
U \\
V \end{array} \right)
&=& \frac{1}{2} \left(
\begin{array}{cccc}
1 & 0 & 0 & 1 \\
\cos 2\psi_{\rm m} & -\sin 2\psi_{\rm m} & -\sin 2\psi_{\rm m} & -\cos 2\psi_{\rm m} \\
\sin 2\psi_{\rm m} &  \cos 2\psi_{\rm m} &  \cos 2\psi_{\rm m} & -\sin 2\psi_{\rm m} \\
0 & -i & i & 0
\end{array} \right) 
\left( \begin{array}{c}
\frac{\left< XX^* \right>}{G_X G^*_X} \\
\frac{\left< XY^* \right>}{G_X G^*_Y} \\
\frac{\left< YX^* \right>}{G_Y G^*_X} \\
\frac{\left< YY^* \right>}{G_Y G^*_Y} \end{array} \right), \label{eqn:StokesMatrix}
\end{eqnarray}
where $\psi_{\rm m}$ is the parallactic angle ($\pm$ elevation angle for Nasmith optics) and $G_X, G_Y$ are the voltage-domain complex gain of the receiving system.
This conversion is carried out with off-line reduction software.

In a practical receiving system, precise polarization measurements are difficult and require careful instrumental calibration. The measurements are degraded by cross-talk, or leakage, between orthogonal polarizations and fluctuations in the complex gain of the receiving system \cite{2001isra.book.....T}. Detailed calibration schemes will be presented in a forthcoming paper.

\section{Verification}\label{sec:verification}
\begin{figure}[ht]
\begin{center}
\includegraphics[width=150mm]{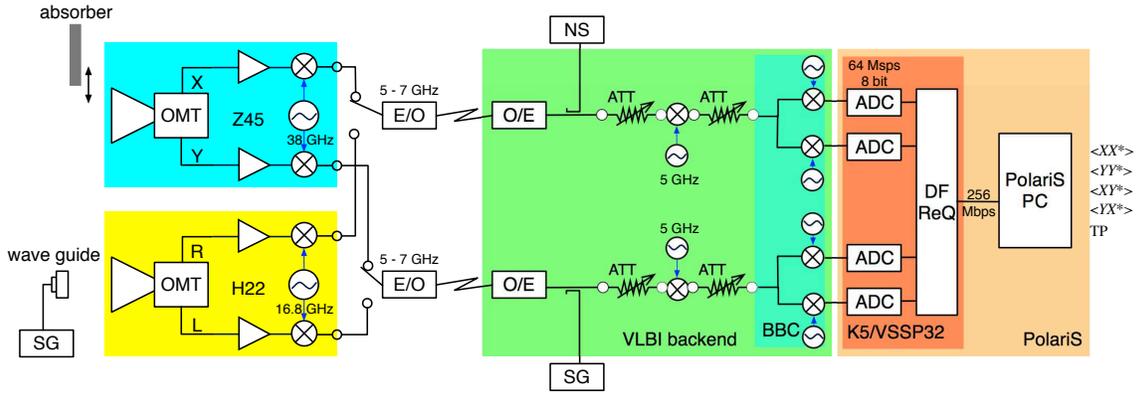} 
\end{center}
\caption{The test configuration.
{ PolariS in pale orange} receives baseband analog signals generated by the
{ BBC} of the VLBI backend. Monochromatic wave generated by the
{ SG} (Agilent E8257D) or white noise produced by the noise source (NTK 9009ZR) were optionally coupled to measure the spectral resolution function and the linearity in line intensity and total power, respectively. Linear polarizations ($X$ and $Y$) and circular polarizations ($R$ and $L$) of the Z45 { (pale blue)} and H22 { (yellow)} receivers, respectively, were selectively transmitted to the VLBI backend { in green}. An absorber was inserted on the feed horn of the Z45 receiver during noise temperature measurements. Linearly polarized monochromatic wave at 22 GHz was injected to the H22 receiver through the feed, generated with a
{ SG} (Agilent 83650L) and a transducer with a rectangular waveguide to verify the cross-correlation capability of
{ PolariS}.} \label{testConfig}
\end{figure}

We have tested the following aspects of
{ performance}: spectral resolution (subsection \ref{sec:sec_resofunc}), linearity (subsection \ref{sec:linearity}), stability (subsection \ref{sec:stablity}), sensitivity (subsection \ref{sec:sens}), and cross-correlation capabilities (subsection \ref{sec:corr}). We used the Z45 receiver and the H22 receiver, working at 45 and 22 GHz, respectively, installed on the NRO 45-m telescope.

\subsection{Spectral resolution function}\label{sec:sec_resofunc}
\begin{figure}[ht]
\begin{center}
\includegraphics[width=3in]{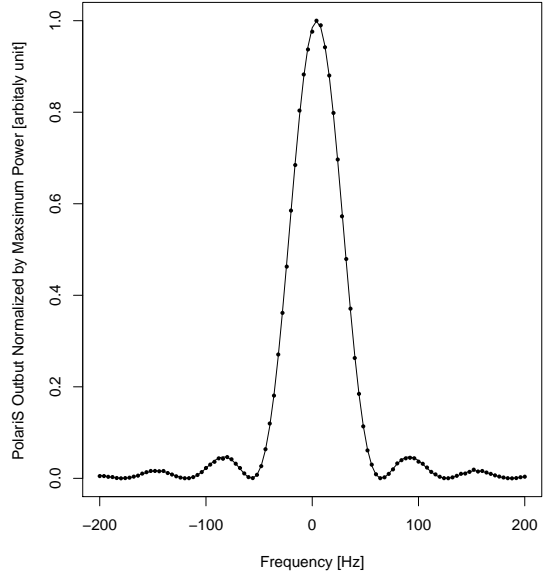} 
\end{center}
\caption{The spectral resolution of
{ PolariS}. Measurements (dots) and the best-fit
{ sinc-squared} function with the FWHM of 54.02 Hz (solid line) are displayed.}
\label{aba:resofunc}
\end{figure}

 The spectral resolution function is the response of the system to a monochromatic input. It demonstrates the sensitivity of the system as a function of frequency and is an indicator of the coupling between adjacent spectral channels (also known as spectral sidelobes). Any spectral feature measured by the spectrometer is the convolution of of the true line shape and this spectral resolution function. The spectral resolution function of PolariS is determined by the FFT process where input data stream is truncated in finite-length segments. PolariS employs a rectangular window function. This makes the spectral response after FFT the
 { sinc} function. Thus, the spectral resolution function formed by the cross-multiplication of these FFT products (see Fig.~\ref{fig:polariS_Inside}) will be a 
{ sinc-squared} function with the FWHM of 0.886 channel (54.07 Hz for 8-MHz bandwidth). A desired apodization window function can be applied offline to reduce the spectral sidelobes. To minimize the information loss through the FFT process, we do not apply apodization in PolariS. We also employed a 50-\% overlap of adjacent FFT segments to save sensitivity for very narrow spectral feature by keeping the number of correlation measurements at longer time lags \cite{2001isra.book.....T}.

We confirmed that the measured spectral resolution function coincided with the theoretically expected one through the following tests, whose configuration is shown in Fig.~\ref{testConfig}.
A monochromatic wave was generated by the signal generator (Agilent E8257D) and was injected to
{ PolariS}.
Its frequency was swept with a 4-Hz step over a 400-Hz range across the reference spectral channel at 68 MHz.
Since the bandwidth of the monochromatic wave was sufficiently narrow, the response to the reference channel as a function of the frequency separation expressed the spectral resolution function.

To characterize the measured spectral resolution function, we removed the continuum noise level determined by line-free channels.
The calibrated power, $P_{\rm ref}$, as a function of frequency separation of the monochromatic wave is shown in Fig.~\ref{aba:resofunc}.
We applied the least-squares fit with the
{ sinc-squared} function as $\displaystyle P_{\rm ref} \sim a_0 \left( \frac{\sin \left(\pi \left(a_1 \nu - a_2\right)\right)}{\pi\left( a_1 \nu - a_2\right)} \right)^2$, where $a_0$, $a_1$, and $a_2$ were free parameters that related to amplitude, resolution, and the center frequency, respectively.
The best-fit
{ sinc-squared} function is displayed in the
{ solid} line in Fig.~\ref{aba:resofunc}.
The results corresponded to the spectral resolution of FWHM $=54.02 \pm 0.06$ Hz, which was consistent with the ideal expectation.

\subsection{Linearity}\label{sec:linearity}
For an ideal spectrometer, the measurement values of spectral power must be proportional to the input power.
The linear relation between input and output over a enough power range is crucial for accurate spectroscopy.

{ We verified the linearity performance in two aspects of measurements: the total power and the spectral line intensity.
The former is required to measure system noise temperature and flux densities of continuum sources, while the latter is necessary to measure accurate line intensities and profiles.}

\subsubsection{Line intensity}
\begin{figure}[ht]
\begin{center}
\includegraphics[width=3in]{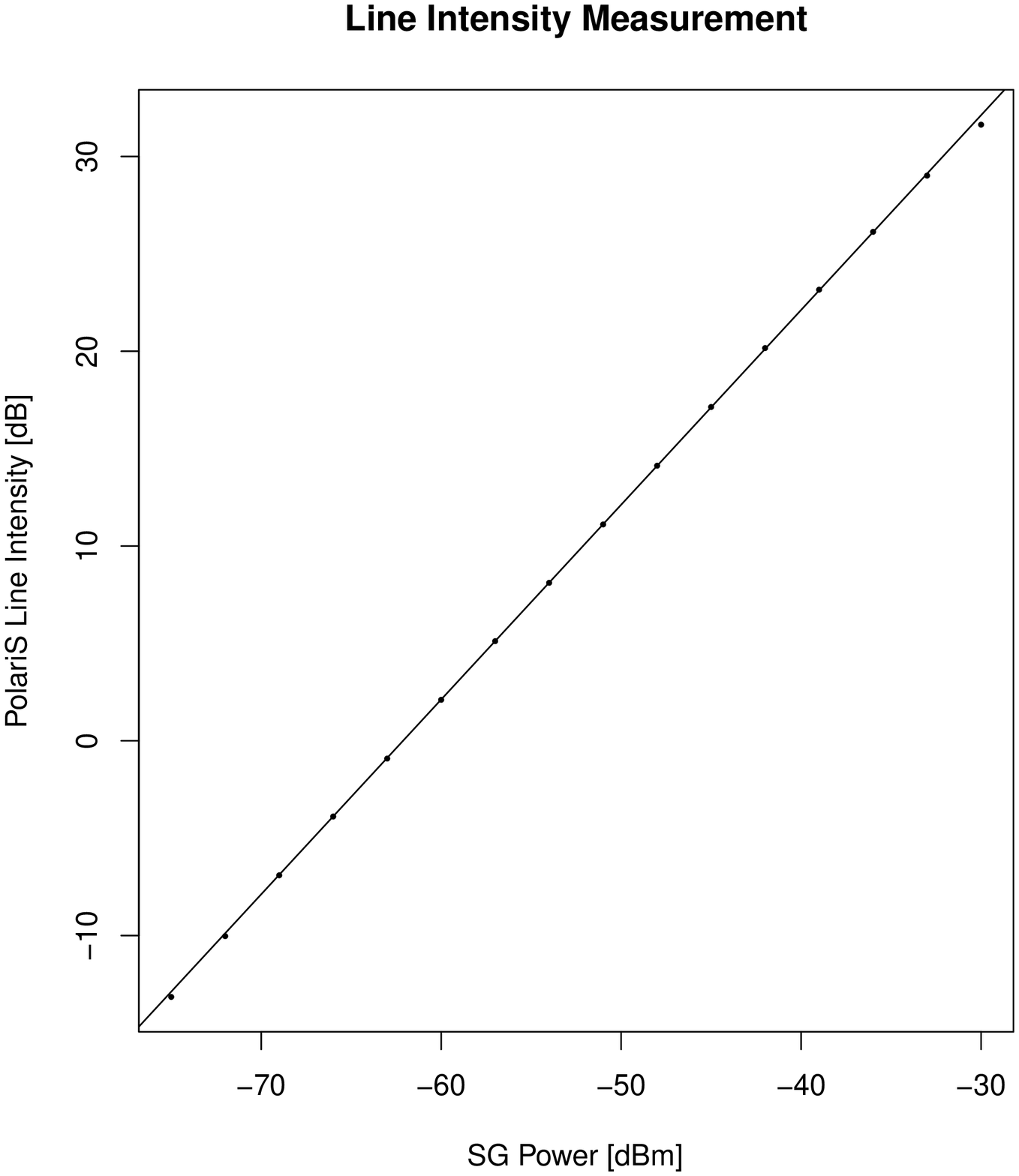}
\includegraphics[width=3in]{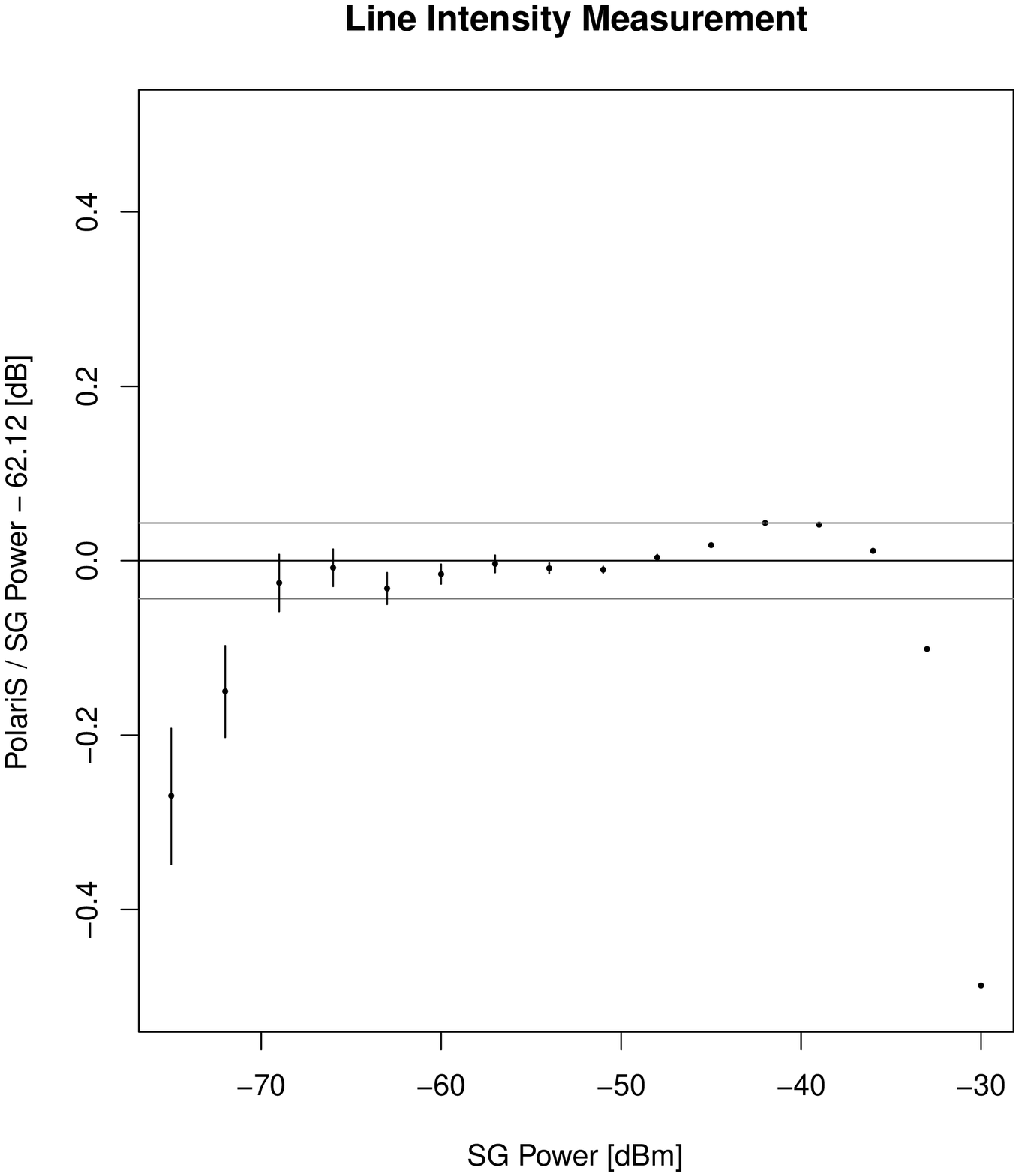}
\\
\includegraphics[width=3in]{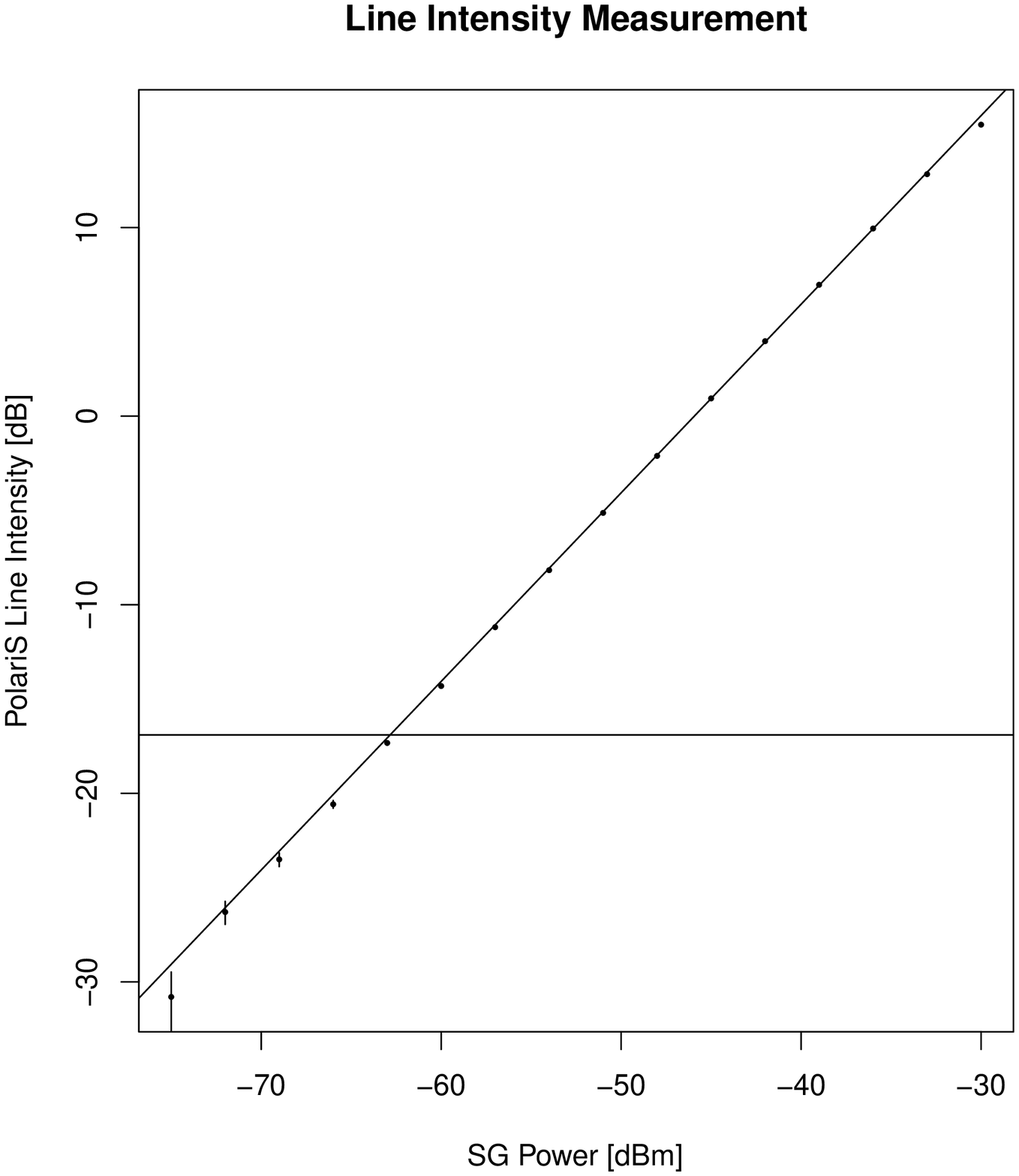}
\includegraphics[width=3in]{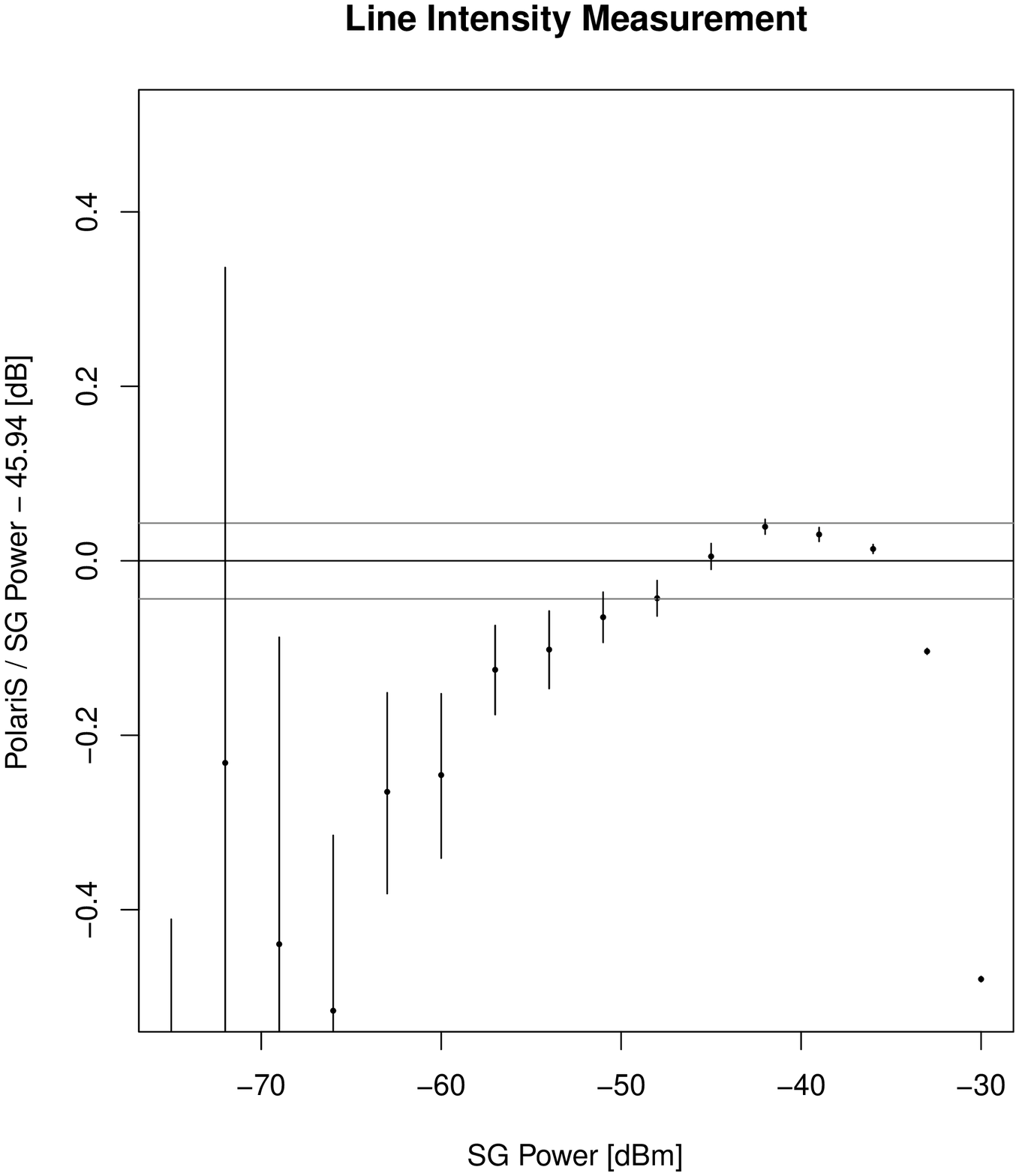}
\end{center}
\caption{{ Linearity in line-intensity measurements.} Top and bottom panels stand for ``high-power'' and ``low-power'' injection of a monochromatic wave produced in the
{ SG}. The low-power injection employs an additional 16-dB attenuator compared with the high-power circumstance.
(Left): Line 
{ intensity} measured by
{ PolariS} as a function of input power (dots).
{ The solid diagonal line indicates the best-fit linear relation with the slope of unity.}
The
{ gray} horizontal line indicates the standard deviation of the system noise level.
(Right): Departure from the 
{ linear relation marked as the solid diagonal line.}
Horizontal
{ gray} lines indicate the
{ departure of $\pm 1$\% range}.
}
\label{aba:linelini}
\end{figure}

We used the Z45 receiver pointing to the blank sky and injected a monochromatic wave generated by the signal generator (Agilent E8257D) into the IF signals (see the configuration in Fig.~\ref{testConfig}) to simulate astronomical spectral-line observations.
The frequency of the injected signal was fixed at 5449 MHz, corresponding to 1 MHz in
{ the baseband}, and the signal power was changed between $-30$ dBm and $-75$ dBm with a 3-dB step.
To cover a wider power range, we used two additional attenuators of 20 dB (`high power') and 36 dB (`low power') for the
{ injection}.
We set
{ PolariS} at the 4-MHz bandwidth, 8-bit quantization and 65536-channel mode. To measure the line power, we integrated the acquired spectrum for 60 sec at each SG power level and subtracted the continuum baseline 
{ as we did for the spectral resolution function (section \ref{sec:sec_resofunc}).}
The line power at the reference channel ($P_{\rm ref}$) was compared with the power of injected monochromatic wave, $P_{\rm in}$.

Fig.~\ref{aba:linelini} shows the results, together with the linear regression of $P_{\rm ref} = b P_{sg}$, where $b$ is the proportional coefficient. 
The proportional coefficients at `high-power' and `low-power' circumstances were 62.12 dB and 45.94 dB, respectively.
The residuals from the linear regression, $\displaystyle \frac{P_{\rm ref} - b P_{\rm in}}{b P_{\rm in}}$, are also displayed in Fig.~\ref{aba:linelini}.
The results indicated that the departure from the linear regression was less than 1\% within the power range of 33 dB and 12 dB in `high-power' and `low-power' circumstances, respectively.

\subsubsection{Total power}
\begin{figure}[ht]
\begin{center}
\includegraphics[width=3in]{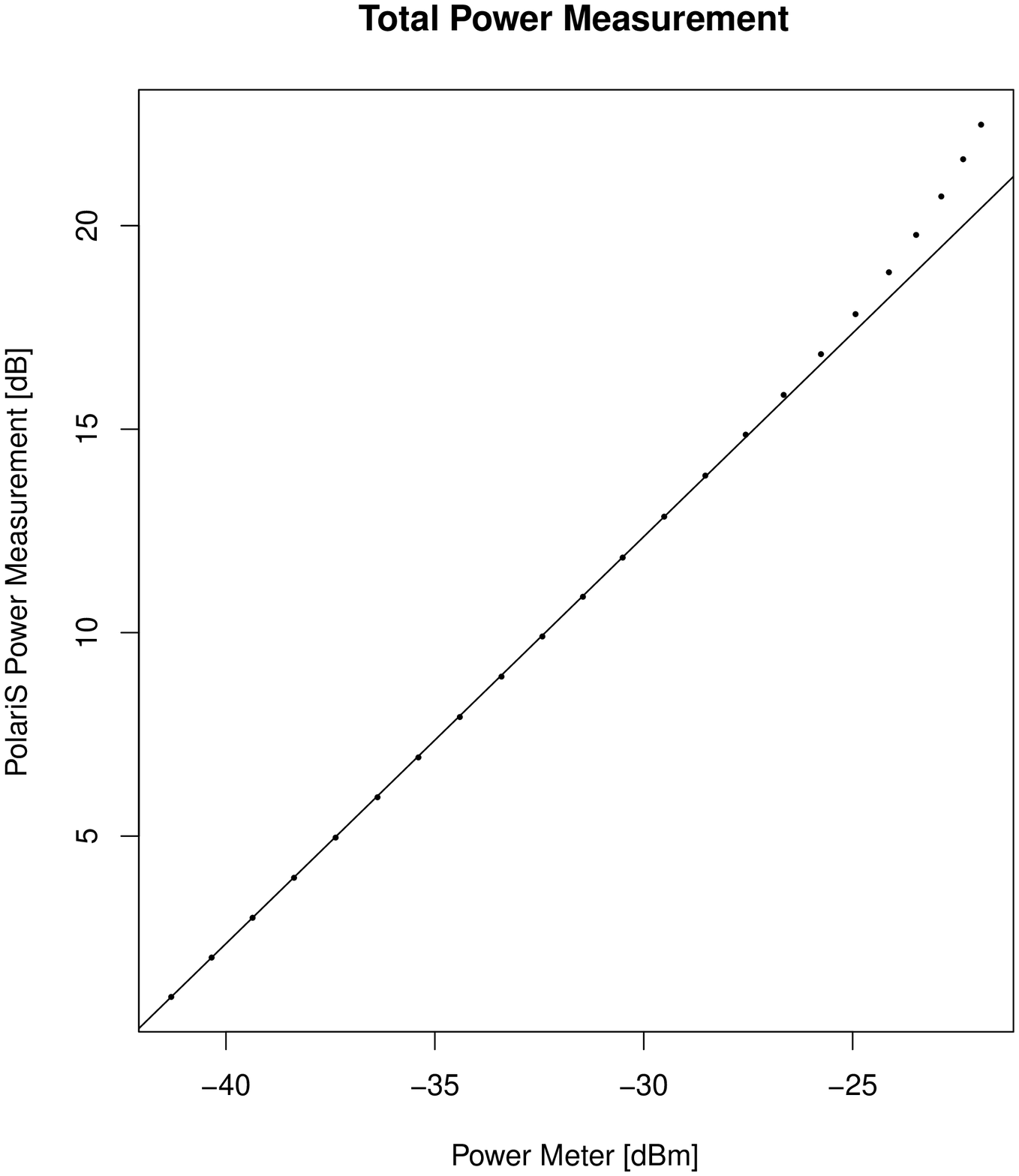}
\includegraphics[width=3in]{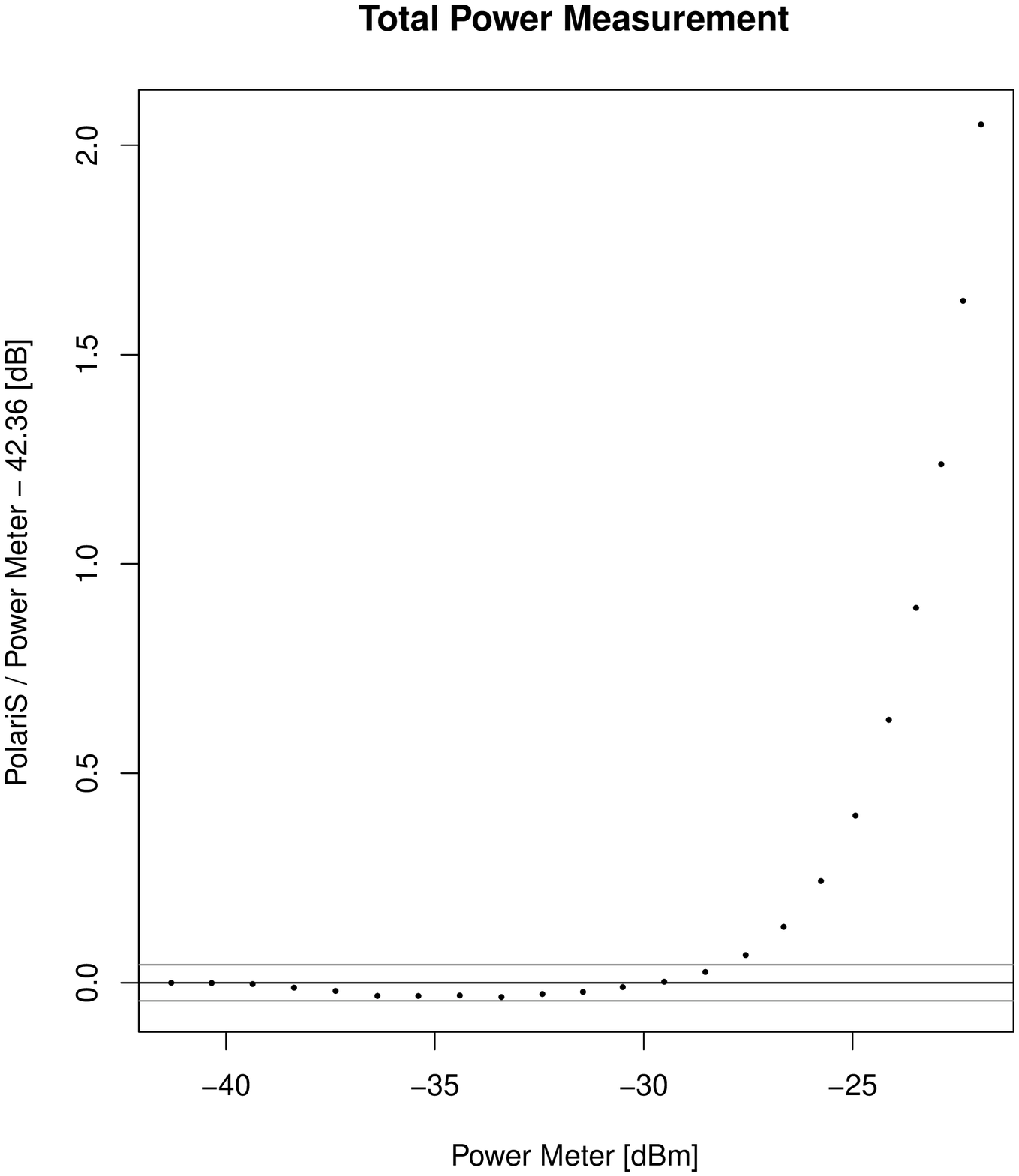}
\end{center}
\caption{(Left)
{ Total power measurements by PolariS} as a function of input power.
{ The solid diagonal line indicate the best-fit linear relation with the slope of unity.}
(Right) Departure from the linear
{ relation}.
{ The gray} horizontal lines indicate
{ departure of $\pm 1$ \% range}.
}
\label{aba:Contlini}
\end{figure}

To verify the linearity of total power measurements, we injected the white noise generated by the noise source (NTK 9009ZR) and adjusted the power level using a combination of two step attenuators (ATTs) covering a 22-dB range by a 1-dB step. 
We employed the power measurement scheme using the bit distribution of digitized signals in the analog-to-digital converter \cite{2010PASJ...62.1361N}.
The input analog power was monitored using the power meter (HP E4419A) with the power sensor (HP 8481D).
The test configuration is shown in Fig.~\ref{testConfig}, { too}.

The
results are shown in Fig.~\ref{aba:Contlini} together with the linear regression of $\log P_{\rm PolariS} \sim \log P_{\rm in} + c$, where $c$ is the sensitivity coefficient that corresponded to 42.46 dB.
Relative departure from the linear regression, $\displaystyle \frac{\log P_{\rm PolariS} - (\log P_{\rm in} + c)}{\log P_{\rm in} + c}$, was within 1\% in the range of $>13$ dB. 

\subsection{Stability}\label{sec:stablity}
Time stability 
{ of the transfer function of the system} 
is required for accurate measurements in radio astronomy observations that take a long integration time.
Instability of a spectrometer can cause systematic errors when scans for calibrator and target sources are not simultaneous.
In the case of radio spectroscopy, a spectrum acquired toward the blank sky (off-source) is used to calibrate the bandpass shape, 
{ $H(\nu)$},
and to subtract the system noise \cite{yamaki2012}.
Stability of the bandpass shape is essential for weak-spectral-line detectability.

To evaluate the bandpass stability, we measured the time-based Allan variance (TAV) and estimated the stable timescale.
We employed the SBC (Smoothed Bandpass Calibration) method \cite{yamaki2012} to distinguish the systematic bandpass variation from random noise.
The spectral-smoothing window was determined by the spectral flatness evaluated using the spectral Allan variance (SAV).

To evaluate the TAV and SAV, we conducted spectroscopy of the noise-source signals for 54000 s.
The configuration is { again} shown in Fig.~\ref{testConfig}. 
Note that the measured instability involves not only PolariS but also the whole system.

\subsubsection{Time-based Allan variance (TAV)}
\begin{figure}[htbp]
\begin{center}
\includegraphics[width=3in]{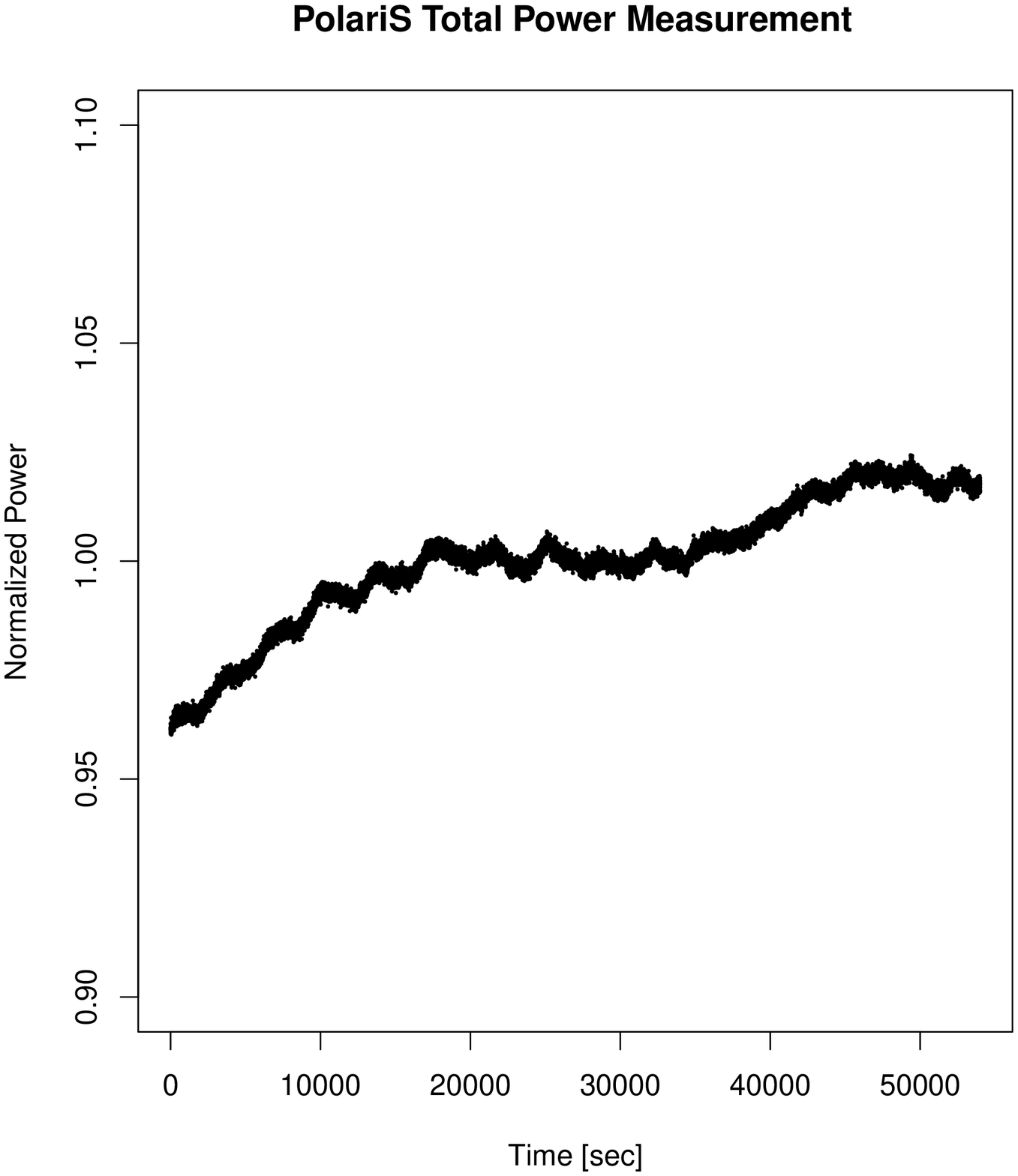}
\includegraphics[width=3in]{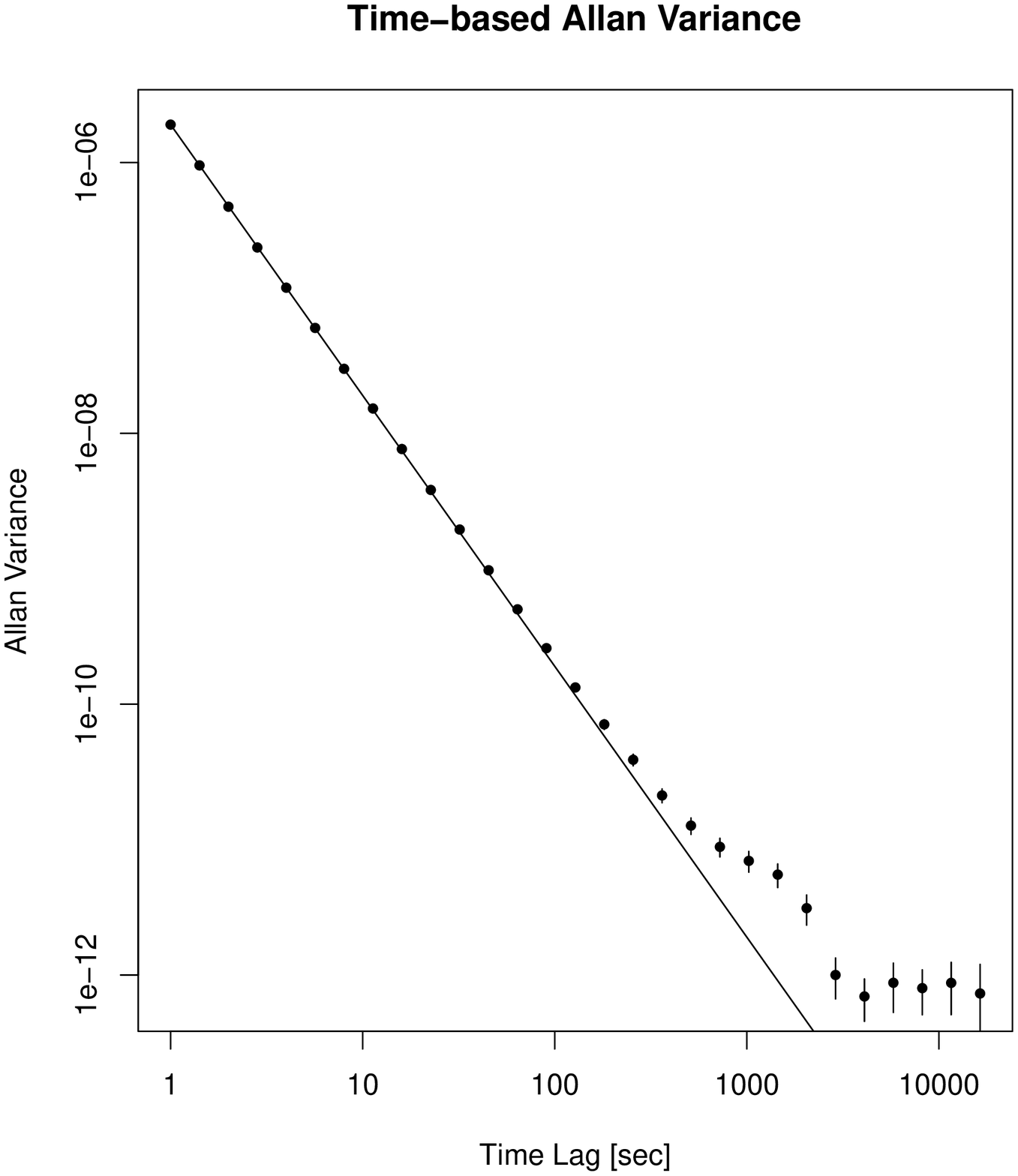}
\caption{(Left): Time variation of the total power measured by the histogram of digitized signals for 54000 s with a 1-s time resolution. The total power was scaled by the mean value. (Right): Time-based Allan variance (TAV) of the total power measurement. The
{ solid} line indicates the power law with the power index of $-2$, expected for white noise. The departure of TAV from the power law became significant at the time lag $\geq 512$ s.}
\label{aba:tavplot}
\end{center}
\end{figure}

For the evaluation of time stability, we calculated the TAV, $\sigma^2_{\rm y}(\tau)$ at the reference spectral channel near the band center, as a function of time lag, $\tau$, defined as: 
\begin{eqnarray}
\sigma^2_{\rm y}(\tau) = \frac{\left< [P_{\rm ref}(t+\tau)-2P_{\rm ref}(t)+P_{\rm ref}(t-\tau)]^2 \right>}{2\tau^2}. \label{eqn:tav}
\end{eqnarray}
The TAV will follow $\sigma^2_{\rm y}(\tau) \propto \tau^{-2}$ while the systematic variation is not significant compared with the white random noise.
We defined the timescale of stability when the excess of $\sigma^2_{\rm y}(\tau)$ from the $\propto \tau^{-2}$ relation is identical to the expectation.

Fig.~\ref{aba:tavplot} shows the total power variation and the TAV measurement.
 The total power in the first 1000 s gradually increased. The variation can be ascribed to warming up of the noise source.
The TAV followed the $\propto \tau^{-2}$ relation for $\tau \leq 256$ s and the timescale of stability was 512 s.
Note that the TAV included not only variations of PolariS but also the analog devices such as the noise source, the IF downconverter, and the baseband converter.
The stability of solo PolariS must be better than the measured TAV.

\subsubsection{Bandpass stability}
\begin{figure}[htbp]
\begin{center}
\includegraphics[width=6in]{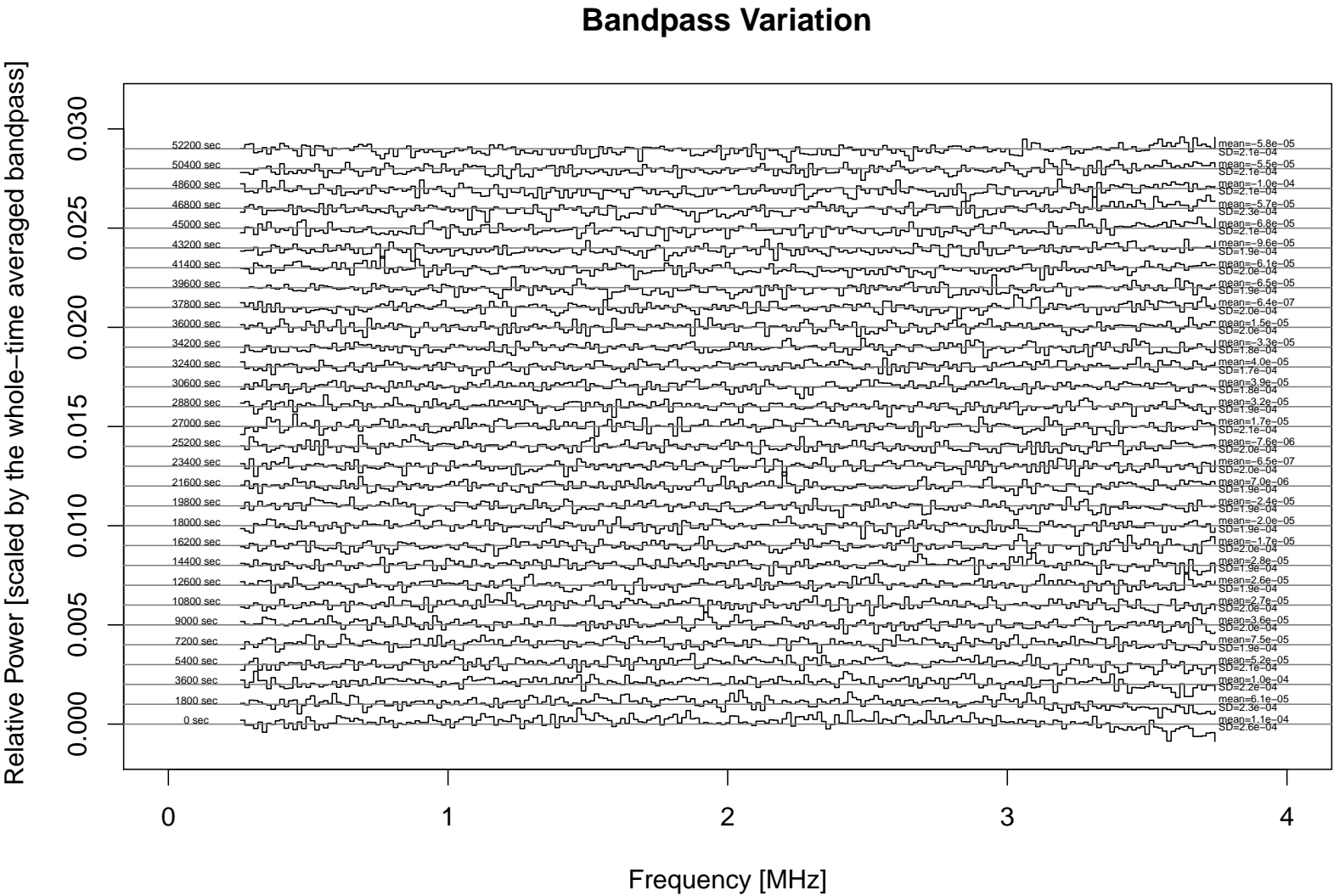}
\caption{
Time variation of bandpass shape.
{ The series of bandpass-corrected spectra, $\hat{H}_k(\nu) \ (k = 1 \dots 30)$, are plotted with the { offset} baselines by the gray horizontal lines}.
The scale of the vertical axis is relative to the 54000-s averaged bandpass, $H(\nu)$.
Relative time from the start is labeled at the left of each spectrum, together with the mean and SD values at the right.
}
\label{BPvar}
\end{center}
\end{figure}

\begin{figure}[htbp]
\begin{center}
\includegraphics[width=3in]{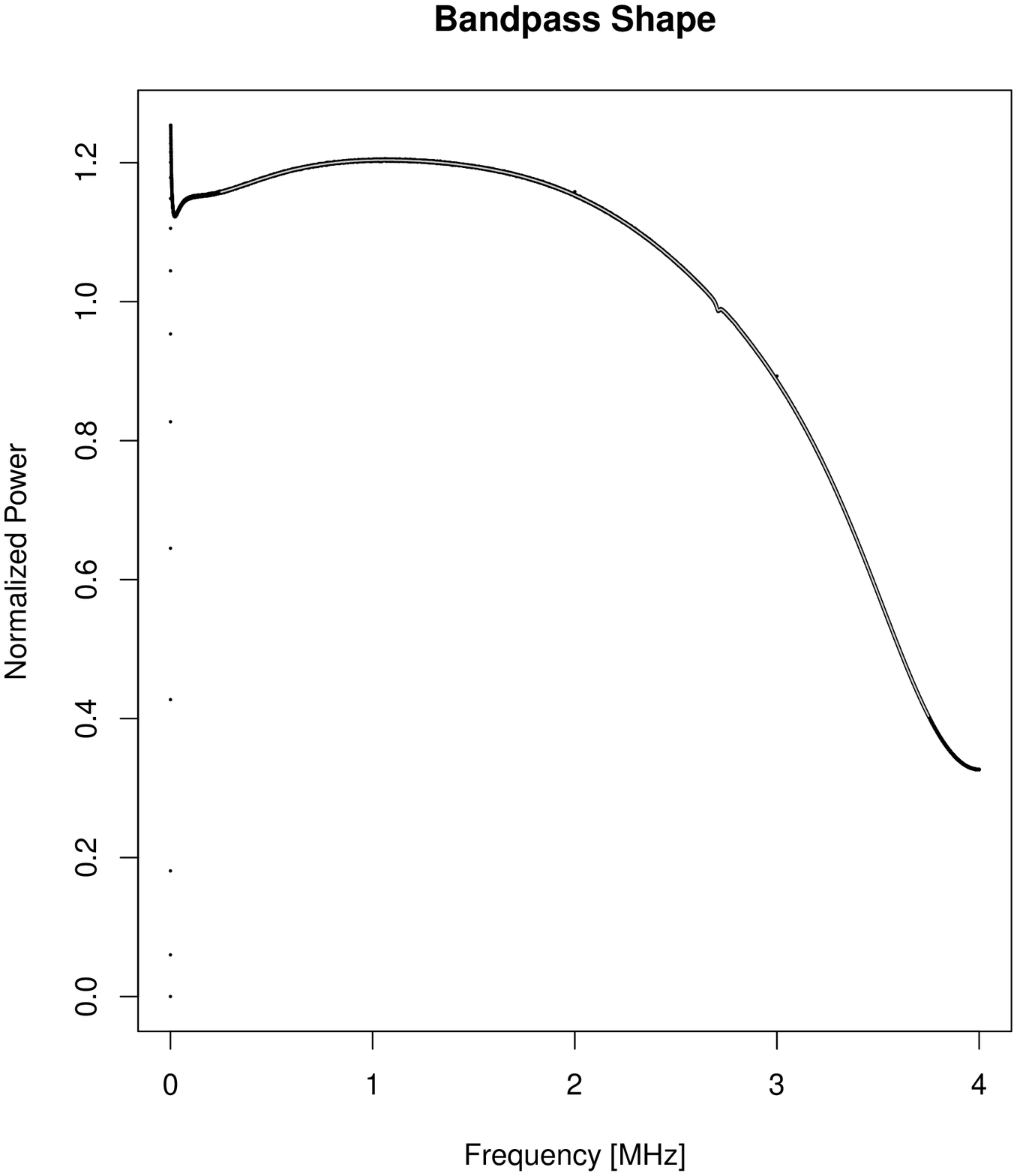}
\includegraphics[width=3in]{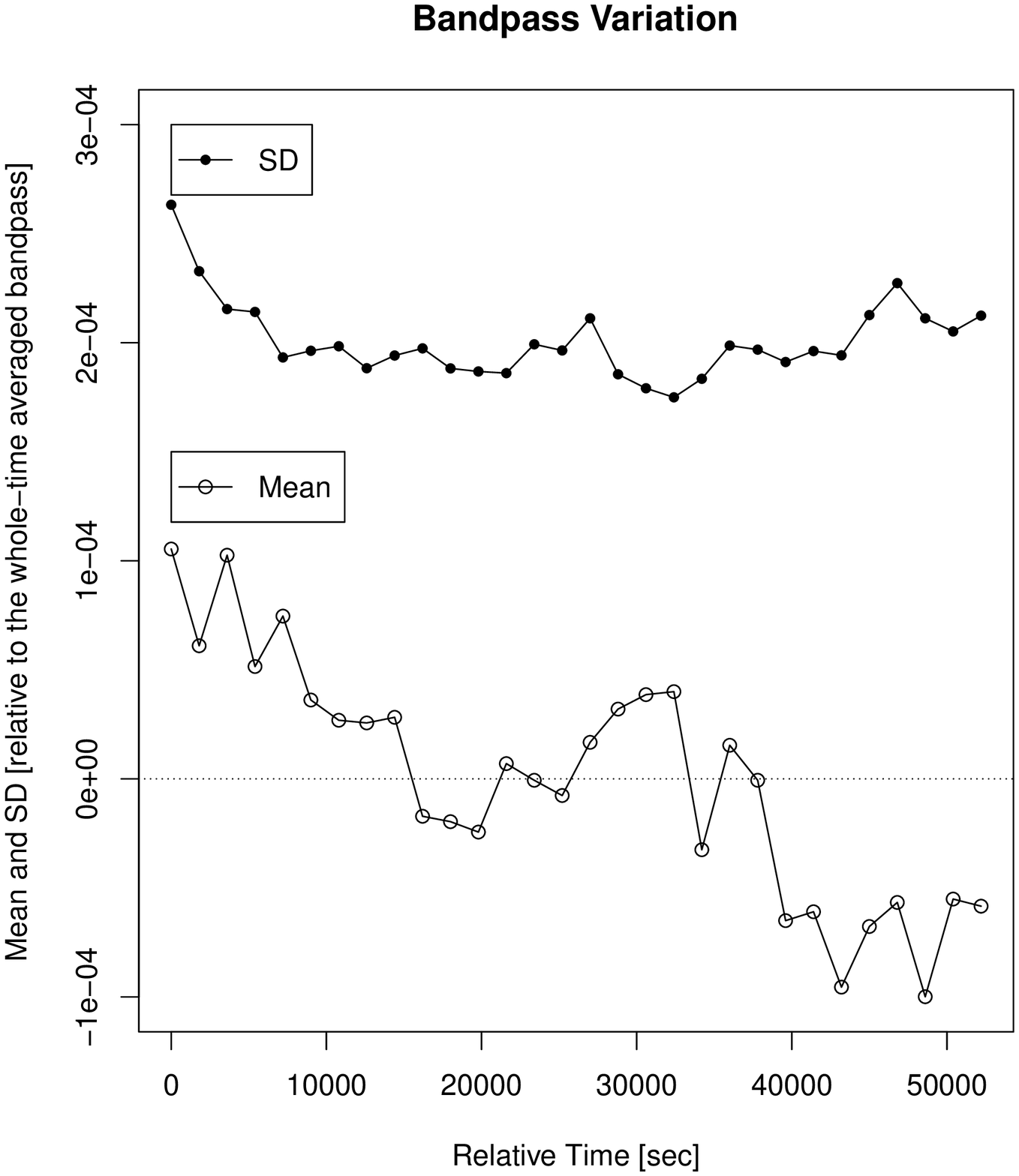}
\caption{
 (Left): Whole-bandwidth spectrum integrated for 54000 s. The
 { gray} solid line indicates spline-smoothed bandpass shape without 6\% band edges at both ends. 
(Right): Mean and SD values of calibrated spectra shown in figure 8, as a function of relative time from the start.
}

\label{BPmodel}
\end{center}
\end{figure}

%
%

 
The stability of bandpass shape was examined with the same configuration of the TAV, shown in Fig.\ref{testConfig}.
The overall bandpass table, { $H(\nu)$ was obtained by integrating the spectrum for 54000 s and then applying spline smoothing with a 128-ch width except for 6\% band edges at both ends} as shown { the gray line} in Fig.~\ref{BPmodel}.
{ Time series of 1800-s integrated bandpass variation, $\displaystyle \hat{H}_k(\nu) = \frac{H_k(\nu)}{G_k H(\nu)}$, were shown in Fig.~\ref{BPvar} to evaluate the bandpass stability.}
{ Here, $k = 1 \dots 30$ is the index of 1800-s integration periods, and $G_k$ is the gain table obtained by 1800-s integration of the total power, $P_{\rm PolariS}$.}
{ To reduce random error in the bandpass variation, we averaged $\hat{H}_k(\nu)$ for 256 spectral channels (15.625 kHz).
The mean and standard deviation (SD) values of each bandpass variation are attached in Fig.~\ref{BPvar} and their time variation are plotted in Fig.~\ref{BPmodel}.}
{ We did not find any significant systematic variation of the bandpass shape which exceeded the random-noise SD which was expected as $\displaystyle \frac{1}{\sqrt{\Delta \nu T}} = 2.6 \times 10^{-4}$. }

{ We found a decline of the mean value} through the entire period in Fig.~\ref{BPmodel}, { and will discuss about it} in section \ref{sec:discussion}.

\subsubsection{Spectral Allan variance (SAV)} 
\begin{figure}[htbp]
\begin{center}
\includegraphics[width=3in]{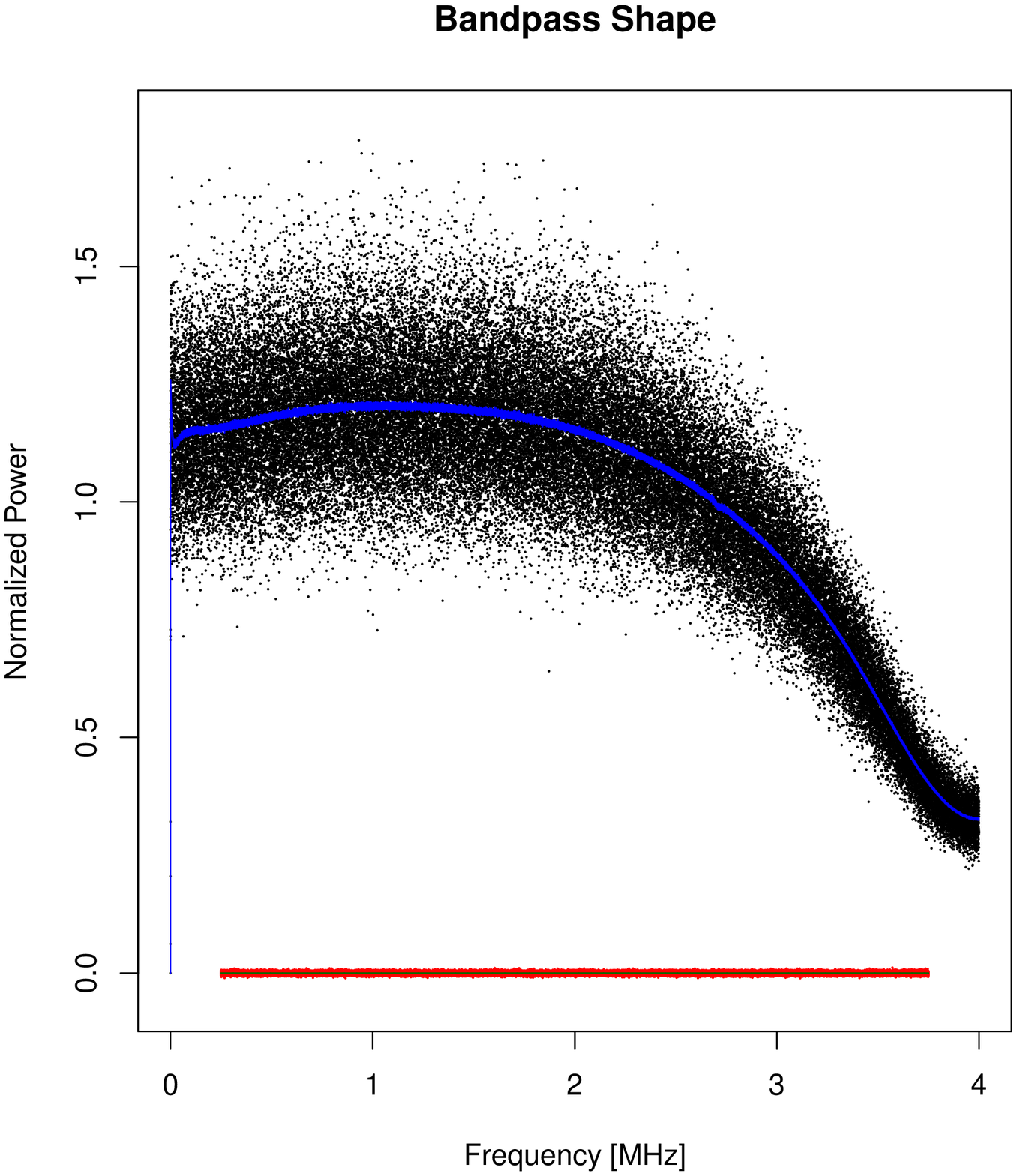}
\includegraphics[width=3in]{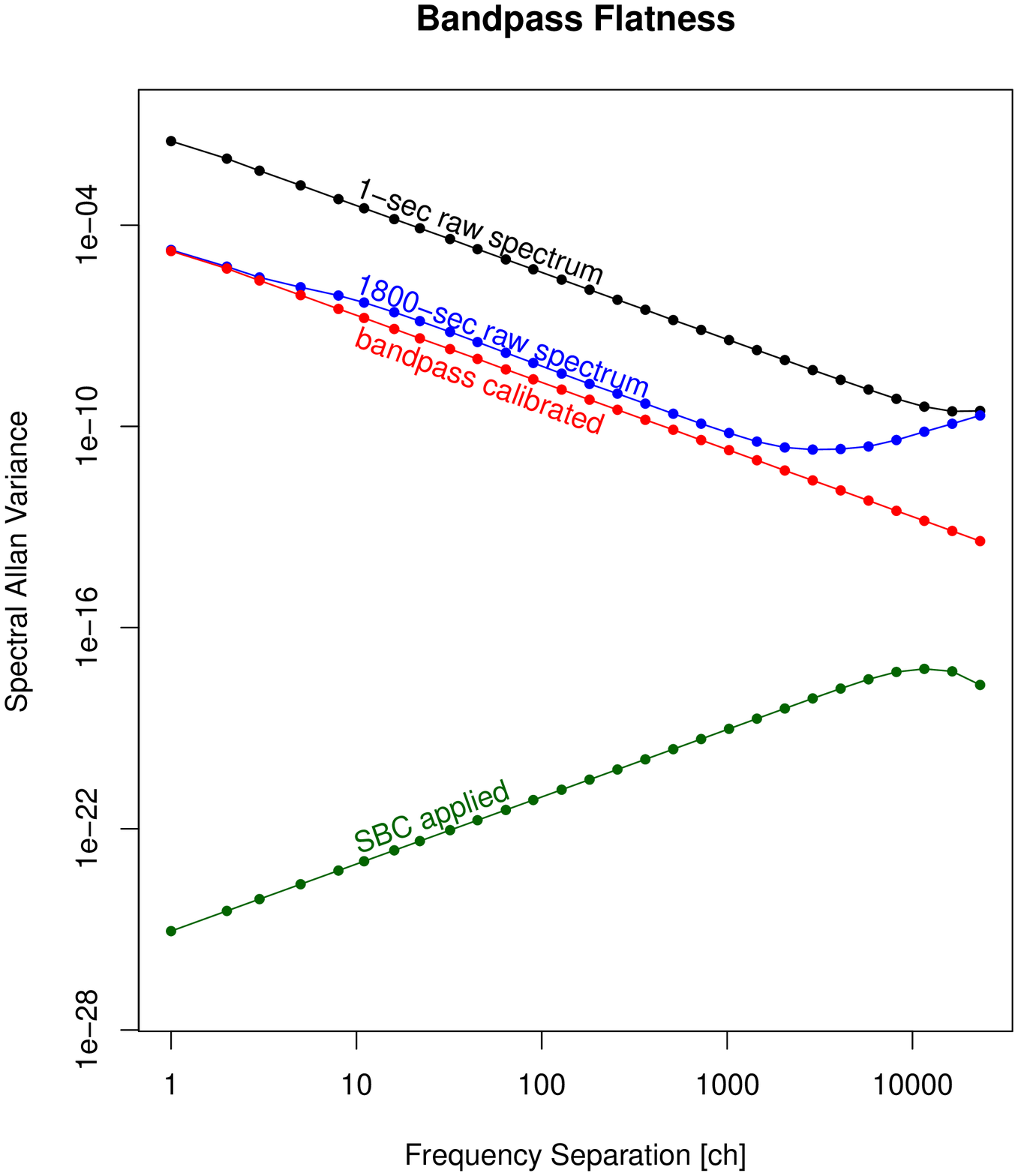}
\caption{(Left): { Spectra at steps of bandpass calibrations}. Black and blue dots indicate the raw spectra before and after 1800-s integration, respectively. The red dots indicate the bandpass-calibrated spectrum. { The SBC-applied spectrum} is plotted in the green line. (Right): Spectral Allan variances (SAVs). Color indices are the same { situation} with the spectra in the left panel.
}
\label{savplot}
\end{center}
\end{figure}

The spectral Allan variance (SAV) defined by equation \ref{eqn:sav} is an indicator of bandpass flatness as a function of frequency separation, $\Delta \nu$.
\begin{eqnarray}
\sigma^2_{\rm y}(\Delta \nu) = \frac{\left< [H(\nu+\Delta\nu)-2H(\nu)+H(\nu-\Delta\nu)]^2 \right>}{2\Delta\nu^2}. \label{eqn:sav}
\end{eqnarray}
While a spectrum is dominated by white noise, the SAV is given by the SD of bandpass-corrected spectrum, $\sigma = \sqrt{\left< H^2(\nu)\right> - \left< H(\nu) \right>^2}$, and the frequency separation as $\displaystyle \sigma^2_{\rm y}(\Delta \nu) = \frac{3 \sigma^2}{\Delta\nu^2}$ and it follows  the power law of the frequency separation with the power index of $-2$.
When systematic distortion of the bandpass shape exceeds the white noise, on the other hand, the SAV shows 
 a greater power index
 than $-2$.

Fig.~\ref{savplot} (right) shows the { measurements of SAV}, together with corresponding bandpass shapes in the Left.
{ The SAV} with a short (1 s) integration was dominated by white noise { that yields the power index of $-2$ in SAV}.
{ Time integration for 1800-s reduced the overall SAV, and it made the bandpass undulation apparent} 
at $\Delta \nu \geq 4$ ch and became dominant at $\Delta \nu \geq 2048$ ch.
The systematic undulation component eliminated from the SAV after bandpass calibration was applied.
The random noise component in the bandpass-calibrated spectrum was efficiently reduced by applying SBC.

\subsection{Sensitivity}\label{sec:sens}
The sensitivity of a spectrometer is 
 determined 
 by the standard deviation (SD) of line-free spectral channels after bandpass calibration and system-noise subtraction because detection of emission (or absorption) lines is judged by the signal-to-noise ratio (SNR), which is the ratio of line intensity to the SD.
When the SD is dominated by random noise, the SD will be reduced by long time integration as $\sigma (T) \propto T^{-1/2}$, where $\sigma$ is the SD with respect to the system noise and $T$ is the integration time.

To evaluate the sensitivity, we used the same dataset for the SAV, bandpass stability, and TAV evaluations and measured the SD using
{ PolariS} for time integration up to 54000 s.
The bandpass calibration table was generated from the spectrum integrated for 1800 s and applied the SBC.
The SD of the spectrum was calculated in the frequency range of 1 -- 2 MHz where the bandpass shape was flat and no significant spurious signals were detected.

\begin{figure}[ht]
\begin{center}
\includegraphics[width=3in]{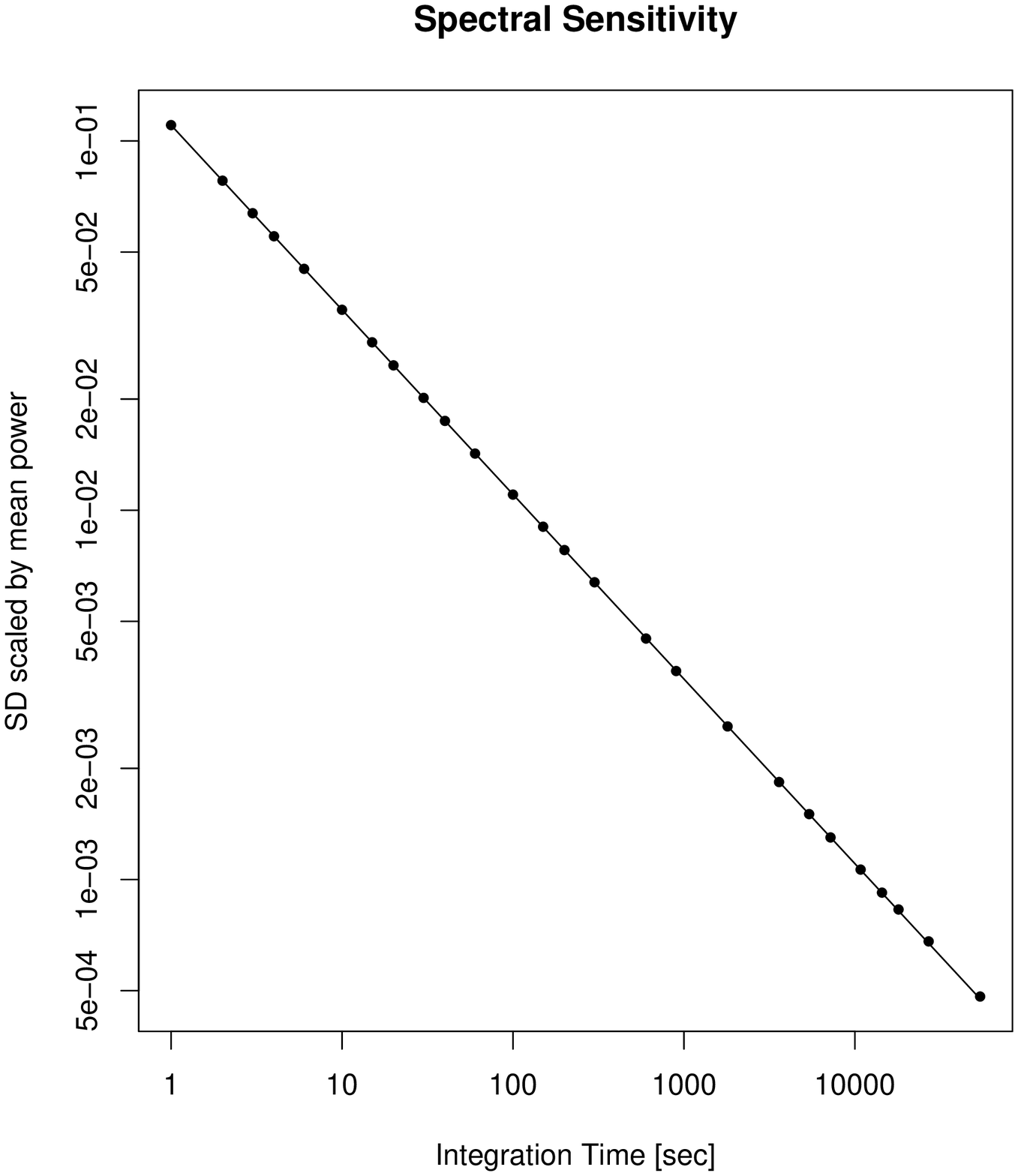}
\includegraphics[width=3in]{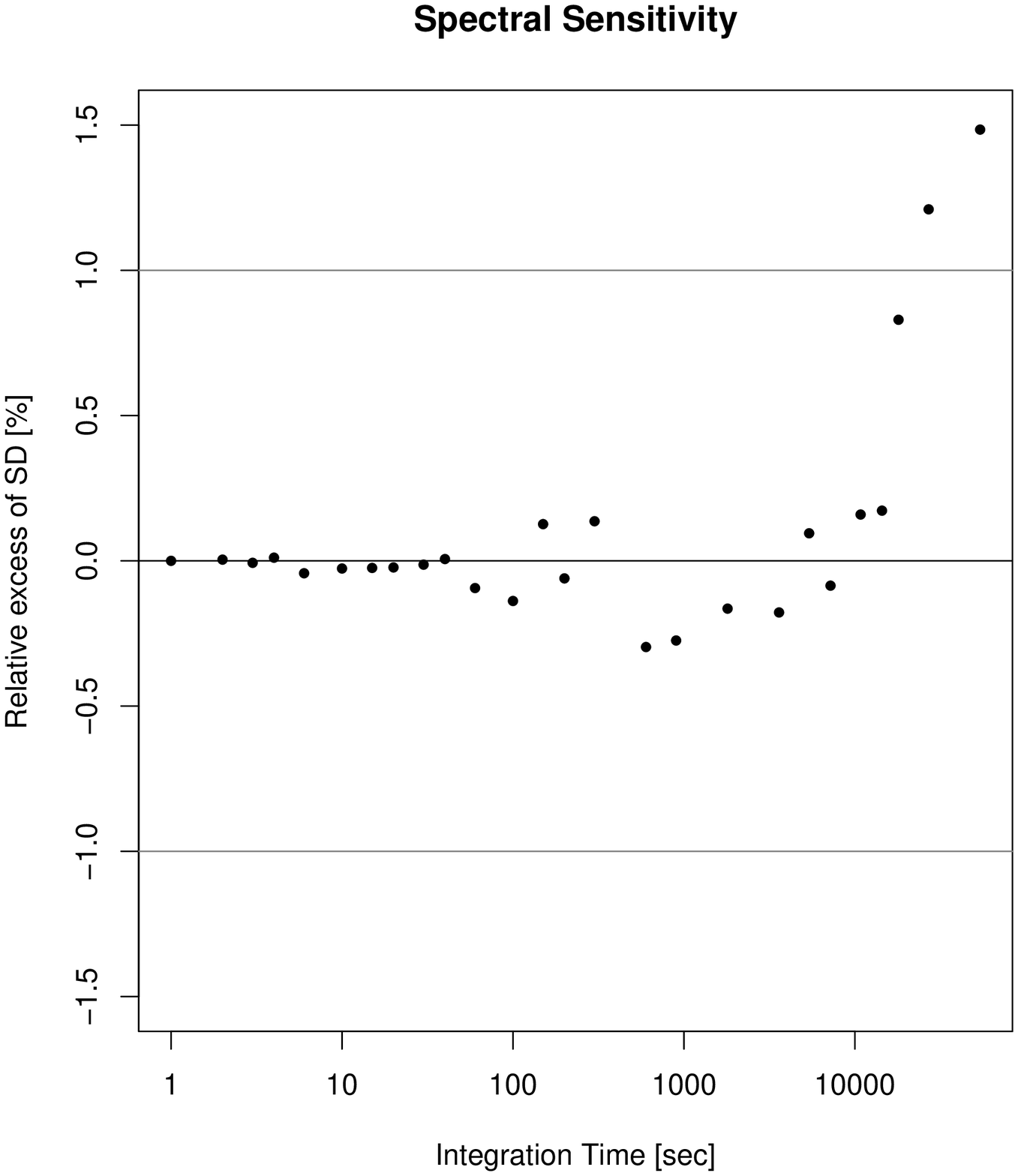}
\end{center}
\caption{(Left): The root-mean-square (RMS) of bandpass-corrected spectra in the frequency range of 1 -- 2 MHz with various integration time, $T$, ranged 1 -- 54000 s.
The 
{ solid diagonal} line indicates extrapolation from the RMS at 1 s with the $\propto 1/\sqrt{T}$ law.
(Right): Relative excess of the RMS from the $\propto 1/\sqrt{T}$ law. The
{ gray} horizontal lines indicate $\pm 1$\% range from the expectation.}
\label{aba:sd}
\end{figure}

Fig.~\ref{sec:stablity} shows the results of the measurements.
The SD at 1-s integration, scaled by the system noise, was 0.110, which was 13.8\% as small as the theoretical value of $\displaystyle \sigma = \frac{1}{\sqrt{\Delta \nu T}}$.
At 54000-s integration, the SD decreased down to $4.82 \times 10^{-4}$ which was still 12.5\% as small as the theoretical expectation.
If we use the SD at 1 s as the reference (see the right of Fig.~\ref{aba:sd}), the SD follows the $\propto T^{-1/2}$ law up to the integration of 18000 s. within 1\% excess.

\subsection{Cross-correlation}\label{sec:corr}
\begin{figure}[ht]
\begin{center}
\includegraphics[width=3in]{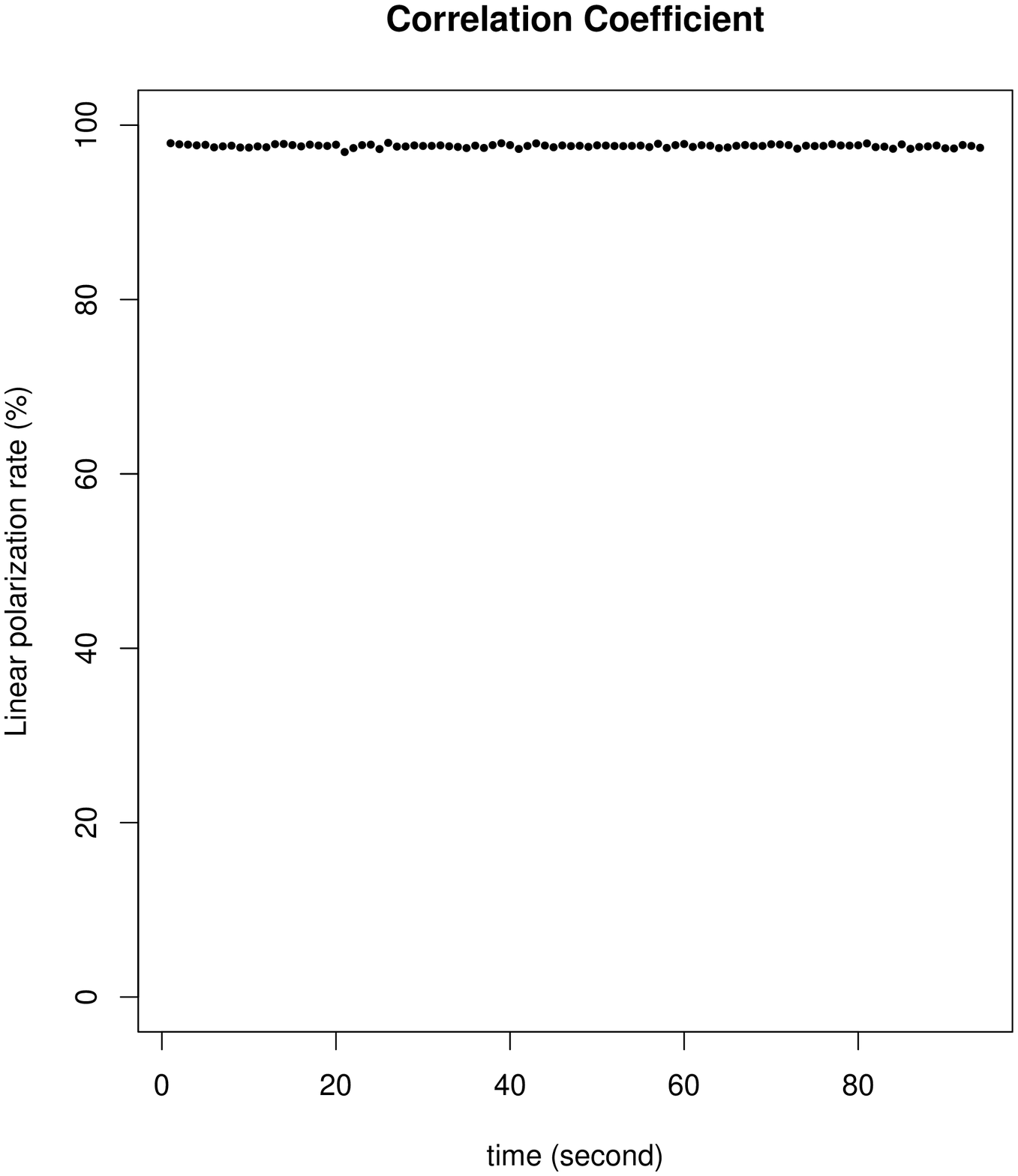}
\includegraphics[width=3in]{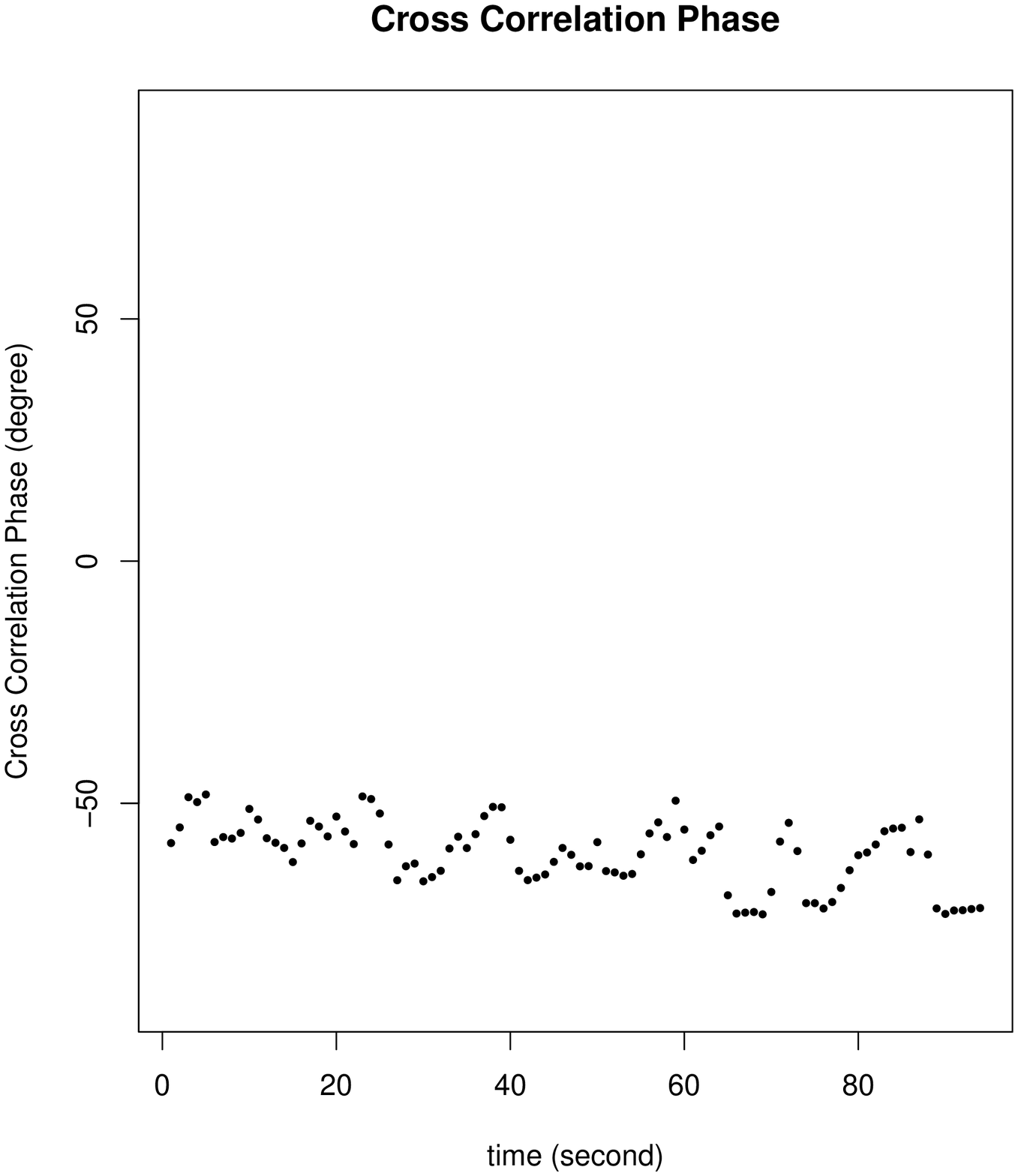}
\end{center}
\caption{(Left): Correlation coefficient between two circular polarization signals taken by
{ PolariS}. The input signal was the linearly polarized monochromatic wave, which was expected to perfectly correlate.
(Right): Phases of the correlation function, which corresponds to relative phase between two circular polarization signals. The standard deviation of the phase was $-60^{\circ}.4 \pm 6^{\circ}.8$.}
\label{aba:polplot}
\end{figure}

The cross-correlation function between two orthogonal polarizations is necessary to obtain full Stokes spectra, as stated in section \ref{sec:stokes}.
To verify the cross products of
{ PolariS}, we used the H22 dual circular polarization receiver and injected a linearly polarized monochromatic wave at 22.233 GHz, 
generated with a signal generator (Agilent 83650L) and a transducer with a rectangular waveguide.
The received two circular polarization signals which  
{ must correlate each other}
were transmitted to the VLBI backend, and acquired with
{ PolariS}. 
The whole configuration is shown in Fig.~\ref{testConfig}, { too}.
The cross-correlation products of
{ PolariS} were tested in terms of the correlation coefficient and relative phase for 94 s with the time resolution of 1 s.
The results are shown in Fig.~\ref{aba:polplot}.
The correlation coefficient was $97.6 \pm 0.2$ \% and the standard deviation of phase was $6.8^{\circ}$ during the observation of 94 s.
We discuss the reason why the correlation differs from 100\% in section \ref{sec:discussion}.

\section{Field Test Observations}\label{sec:fieldtest}
\begin{figure}[ht]
\begin{center}
\includegraphics[width=3in]{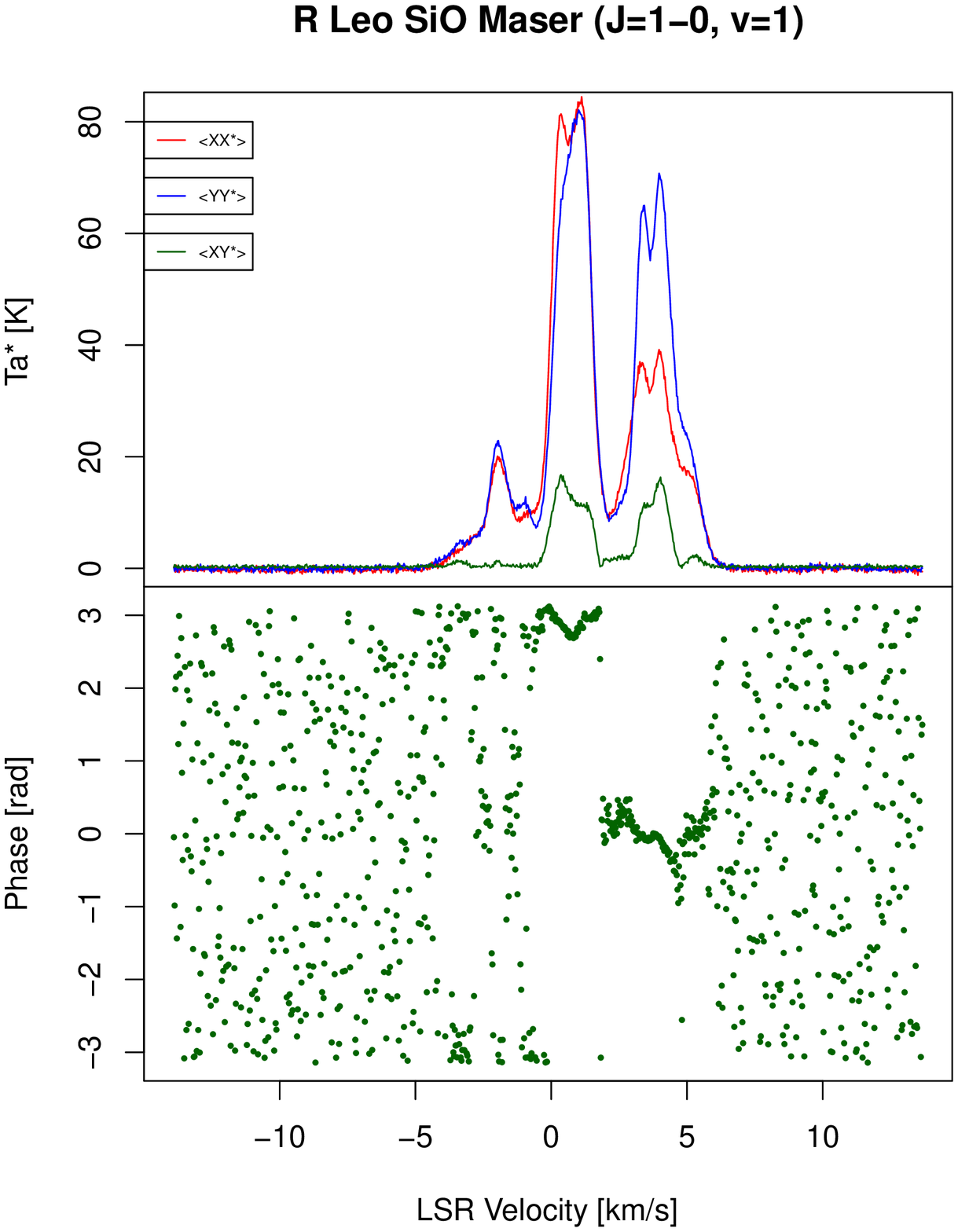}
\includegraphics[width=3in]{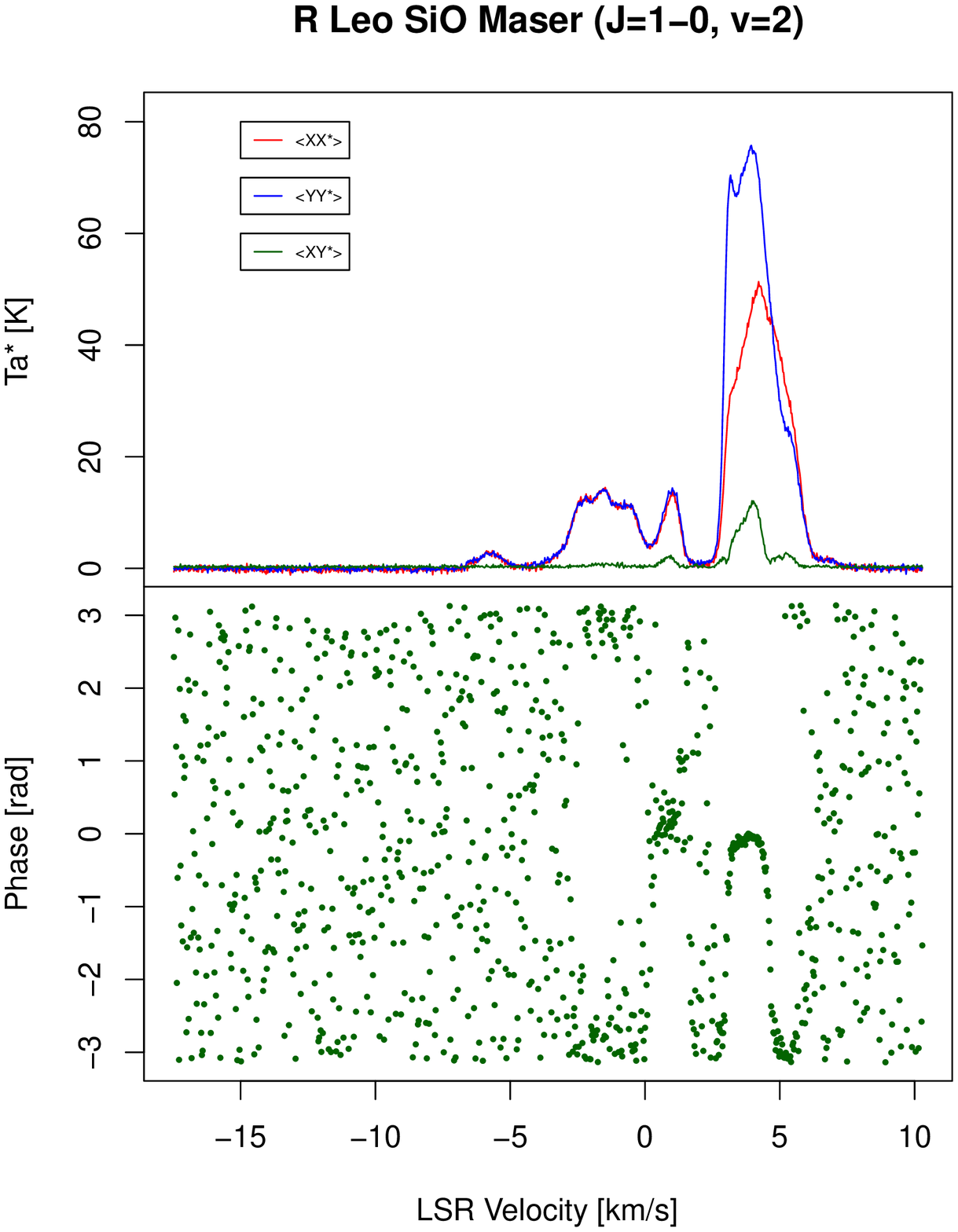}
\end{center}
\caption{Power spectra and cross power spectra of SiO maser ($J=1-0$) emission lines toward R Leo. Left and right panels indicate the vibration state of $v=1$ and $v=2$ at 43.122 and 42.821 GHz, respectively. Red and blue solid lines in the top stand for the power spectra of $X$ and $Y$ linear feeds, respectively. Green solid line and filled circles indicate the amplitude and phase of the cross power spectrum between $X$ and $Y$ polarizations.}
\label{RLeo}
\end{figure}

\begin{figure}[ht]
\begin{center}
\includegraphics[width=3in]{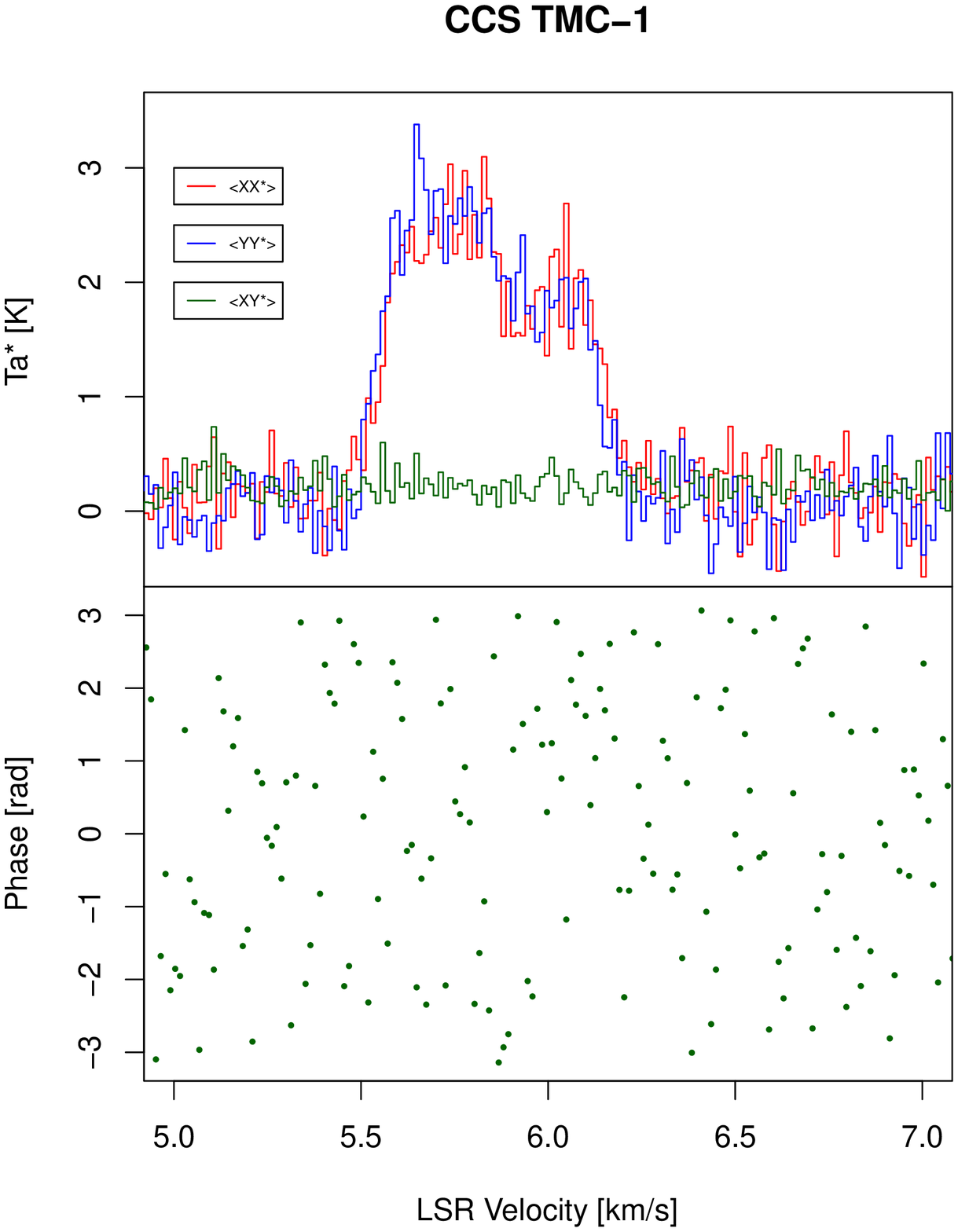}
\end{center}
\caption{Power spectra and cross power spectra of CCS emission line toward TMC-1 at the spectral resolution of 1.95 kHz. Color assignments are the same with Fig.~\ref{RLeo}.}
\label{ccsplot}
\end{figure}

For the purpose of commissioning and science verification, we tried test observations of SiO maser lines toward the Mira variable star, R Leo, and the CCS molecular line toward the star-forming molecular core in the Taurus molecular cloud 1 (TMC-1) using the Z45 receiver on the Nobeyama 45-m radio telescope.

For
R Leo, two transitions of SiO $J=1-0$, $v=1$ (43.122 GHz) and $v=2$ (42.821 GHz) were simultaneously observed with the dual linear polarizations on March 16, 2014.
Spectroscopy was originally carried out with 16384 channels across 4-MHz bandwidth (244-Hz resolution) and
{ channel-averaged} into 3.9-kHz resolution.
The integration time for on- and off-source were 41 s and 65 s, respectively, and the off-source spectra were subtracted from the on-source ones.
The power spectra and cross power spectra are shown in Fig.~\ref{RLeo}.
The power spectra of $\left< XX^*\right>$ and $\left< YY^*\right>$ were significantly different and the cross power spectra were significantly detected.

For the TMC-1, the CCS $J_N=4_3 - 3_2$ transition at 45.379 GHz was observed on December 23, 2013.
We pointed to (RA, DEC)$_{\rm J2000}=(04^h 41^m 42^s.47, \ +25^{\circ} 41^{\prime} 27^{\prime \prime}.1)$, which was close to `b7' clump of TMC-1 core D region \cite{1998ApJ...497..842P}.
Spectroscopy with integration of 252 s and 375 s for on- and off-source was carried out with 131072 channels across 8-MHz bandwidth (61-Hz resolution) and
{ channel-averaged} into 1.95-kHz resolution.
Since a relatively strong spurious signal appeared in one spectral channel near the center of CCS line profile, we flagged the channel before
{ spectral averaging}.
The power spectra and cross power spectra are shown in Fig.~\ref{ccsplot}.

\section{Discussion}\label{sec:discussion}
The engineering tests of
{ PolariS} presented here demonstrate compliance with specifications.

The spectral resolution perfectly coincided with the theoretical expectation, as shown in section \ref{sec:sec_resofunc}.
The current software employs the boxcar window function in the time-domain FFT segments, which results in
{ the sinc-squared} spectral resolution function.
This treatment causes subtle sidelobe associated with narrow and strong emission lines.
PolariS is a software spectrometer, which is flexible for applying other window functions, such as Hamming, Hanning, and Blackman windows, to reduce sidelobe levels in the spectral resolution function.
{ We choose} the boxcar window
{ for online processing} because sidelobe mitigation with a relevant window function can be applied in post processing.
Observers can apply desired apodization function or spectral
{ smoothing}
to reduce spectral sidelobes.

The system
{ marked sufficient} linearity in total power measurements within 1\% { accuracy} over the input power range of $13$ dB.
The total power measurement is used to estimate the receiver noise temperature and the system noise temperature by comparing powers in two or more circumstances where the feed horn points to the absorber at ambient temperature ($\sim 293$ K), to the cold load ($\sim 77$ K), and to the blank sky ($\sim 20$ K).
Since the typical receiver noise of the Z45 is $\sim 50$ K, the power range between the cold load, the blank sky and the ambient load will be $\displaystyle \frac{293 + 50}{77 + 50} = 2.7$ (4.3 dB) and $\displaystyle \frac{293 + 50}{20 + 50} = 4.9$ (6.9 dB).
Thus, the dynamic range of $>13$ dB is sufficient for the system noise measurement, if we set the power level adequately.

We note that the departure from linearity displayed in Fig.~\ref{aba:Contlini} { could be due to saturation of the power meter, because the power meter we employed for this experiment has a maximum rated input power of $-20$ dBm.
Therefore the linear range of PolariS could be wider than presented here.}
 

The dynamic range of 33 dB in line intensity is also enough for astronomical spectroscopy.
The departure from the linear trend in high-power ends can be ascribed to inaccuracy of the signal generator, because the sensitivity profiles above $-40$ dBm are very similar in both `high-power' and `low-power'.
The underestimation in low-power ends, especially in the `low-power' attenuator setting, can be caused by uncertainty in baseline subtraction because the departure from the linear trend was significant where the line intensity is
{ comparable} to the standard deviation ($\displaystyle =\frac{1}{\sqrt{\Delta \nu T}}$) of the system noise.

The performance of weak line detectability is verified with the SD of the spectrum at long integration time.
As is shown in section \ref{sec:sens}, the sensitivity is
maintained for the timescale of 54000 s.
The field-test observations of CCS molecular line toward TMC-1 also indicate that the system shows no significant departure in the linearity range down to $T^*_{\rm a}/T_{\rm sys} = 1.7$\%. 

As shown in subsection \ref{sec:stablity}, the bandpass shape is very smooth and stable, allowing us to apply SBC to reduce {\rm random errors in} off-source scans efficiently.
This indicates that the analog system (baseband converters and analog-to-digital converter) performs sufficiently stably and the digital system (in software) works perfectly.
The SAV (see Fig.~\ref{savplot}) indicates a very good flatness in the bandpass-corrected spectrum.
The systematic error of bandpass shape after SBC application will be less than $\displaystyle \sqrt{\frac{2\Delta \nu^2 {\rm SAV} }{3}} = 2.3 \times 10^{-5}$ at the maximum channel separation of 11585 ch for broad line detection, even if we employ a single bandpass table for 15-hour observations.


The mean values of the bandpass-corrected spectra were decreasing while the increasing total power (see Fig.~\ref{aba:tavplot}).
We consider that {\rm this is ascribed to} increase of power in band edges caused by bandpass distortion. Because the gain table, $G(t)$, is derived from bit distribution that includes power in band edges, increase of power in band edges makes $G(t)$ greater and lets the mean power without band edges decreased after gain calibration.
We can also see convex and concave distortion at the beginning and ending periods, respectively, in Fig.~\ref{BPvar}. This behavior supports the explanation above.

The stability in bandpass shape is also confirmed by the sensitivity shown in subsection \ref{sec:sens}.
As shown in Fig.~\ref{aba:sd}, the SD of bandpass-corrected spectra follows the $\propto T^{-1/2}$ law for the integration time shorter than 18000 s.
The excess of SD at the longest integration time can be caused by the bandpass error, related with variation in total power as discussed in the last paragraph, because the distortion of bandpass shape after SBC is estimated to $\sim 10^{-5}$ and non-negligible for integration longer than 18000 s.
Note that the SD value at the longest integration time is still smaller than the expectation of $\displaystyle \frac{1}{\sqrt{\Delta \nu T}}$.
This is probably because the effective bandwidth is narrower than 4 MHz due to the antialias filter (see Fig.~\ref{savplot}) and thus the sampling frequency of 8 MHz is slightly higher than the Nyquist rate.

The cross-correlation capability is confirmed by the engineering test with the artificial source (subsection \ref{sec:corr}) and the field test (subsection \ref{sec:fieldtest}).
With the injection of the linearly polarized artificial monochromatic wave into a dual circular feed, we got $97.6 \pm 0.2$ \% of correlation.
The departure from $100$ \% could be due to temperature fluctuation of the system that yields relative phase fluctuation between LHCP and RHCP.

The measured cross power spectra of the SiO maser in the observation toward R Leo (see Fig.~\ref{RLeo}) demonstrates the potential of this instrument to perform high precision polarimetric observations.
In both $v=1$ and $v=2$ transitions, the velocity components within $0 \leq V_{\rm LSR} \leq 7$ km s$^{-1}$ shows highly polarized emission with a significant amplitude in cross power spectra, small phase dispersion, and difference between $\left< XX^* \right>$ and $\left< YY^* \right>$ spectra.
The large gap in phase between velocity components of $0 \leq V_{\rm LSR} \leq 2$ km s$^{-1}$ and $2 \leq V_{\rm LSR} \leq 7$ km s$^{-1}$ in $v=1$ suggests that the electric-vector position angles are significantly different between these two components.
The blueshifted components in $-7 \leq V_{\rm LSR} \leq 0$ km s$^{-1}$ show much weaker polarization than the redshifted components.

The CCS molecular emission is expected to be unpolarized, because it is a thermal emission in the dark cloud.
{ As was expected, we got neither significant detection in cross power spectrum nor significant difference between $\left< XX^* \right>$ and $\left< YY^* \right>$ spectra (see Fig.~\ref{ccsplot}).} 
The phases of the cross power spectra randomly distributed in $[-\pi, \pi]$, { suggesting} non-detection.
Note that the amplitude of cross power spectrum is biased from 0, because it is always positive and follows the Rayleigh distribution when the expectation of cross power spectrum is 0.

The line peak intensity of $T^*_{\rm a} =2.57 \pm 0.04$ K with 1.95-kHz resolution is slightly higher than the results of $T^*_{\rm a} = 2.23 \pm 0.09$ K with 37-kHz resolution using the acousto-optical spectrometer (AOS) on the Nobeyama 45-m radio telescope \cite{1992ApJ...392..551S}.
If we apply
{ channel-averaging} to set the same spectral resolution of 37 kHz,
{ the spectrum by PolariS} results in  $T^*_{\rm a} =2.47 \pm 0.04$ K and close to the AOS measurement.
The line profile from the mapping observation of TMC-1 \cite{1998ApJ...497..842P} showed a somewhat different profile, depending on the position of the molecular cloud. Because of the difference of beam size and pointing position, it is difficult to evaluate the difference of line profiles.

The widths of both line edges are only $\sim 20$ Hz. The sharpness of the line profile was $\displaystyle \frac{dI}{d\nu} = 0.16$ mK Hz$^{-1}$ that yields Stokes $V \pm 10$ mK at both line edges if we assume the Zeeman shift of 64 Hz for $B \sim 100 \ \mu$G.
To detect the Stokes $V$ with the signal-to-noise ratio of 5 under the condition of $T_{\rm sys} = 100$ K, on-source integration time of $\displaystyle \left( 5 \times \frac{100 \ {\rm K}}{10 \ {\rm mK}} \right)^2 \frac{1}{2 \times 20 \ {\rm kHz}} = 62500$ s is required.

\section{Summary}
We have developed a software-based polarization spectrometer, PolariS, using { commercially available} devices of the K5/VSSP32
{ sampler} and the Linux computer, at the cost of JPY 500,000 and JPY 70,000, respectively.
Using the GPU for FFT and cross-correlation processes, spectroscopy for 4-IF 4- or 8-MHz bandwidth with a 61-Hz spectral resolution can be carried out in real time.

Through the engineering tests, we have verified the basic performance of
{ PolariS}.
The spectral resolution function was consistent with the theoretical prediction of
{ a sinc-squared} profile with a FWHM of $54.02 \pm 0.06$ Hz.
The linearity between input signal power and measured power has been confirmed within 1\% accuracy for the range of $>13$ dB and 33 dB for continuum total power and emission line, respectively.
The bandpass shape was flat and stable enough to apply SBC with 125-kHz frequency width.
No significant bandpass variation exceeding $2.6 \times 10^{-4}$ with respect to the system noise was found for the time span of 15 hours.
The standard deviation of the spectrum was 12.5\% as small as the expectation of $\displaystyle \frac{1}{\sqrt{\Delta \nu T}}$ at 54000-s integration and follows the $\propto T^{-1/2}$ law between 1 s and 54000 s.
The cross power spectra were verified by injection of a linearly polarized artificial signal into a dual circular feed and cross-correlation of them.

The field test observations were carried out toward the strongly polarized maser source, R Leo, and unpolarized CCS emission of TMC-1.
We verified the performance of the power spectra that will yield full Stokes spectra crucial to our primary aim of the Zeeman effect measurements.

All
{ the software code} is open in the GitHub repository and anyone can download and use it or review the signal processing inside.

\section*{\it Acknowledgments}
The development of
{ PolariS} is supported by the FY2012 Joint Development Research program of the National Astronomical Observatory of Japan, and by a Grant-in-Aid for Scientific Research of Japan (24244017).
We thank the staff and students in the Department of Physics and Astronomy, Kagoshima University for assistance with fundamental tests using the VERA Iriki station. We are grateful to the staff of the Mizusawa VLBI Observatory for technical support and commissioning. The authors thank the staff at the Nobeyama Radio Observatory (NRO) for operations of the 45-m radio telescope in the commissioning science verification. NRO is a division of the National Astronomical Observatory, National Institutes of Natural Sciences, Japan.

\end{document}